\documentclass[12pt,preprint]{aastex}
\newcommand{\degree}{$^{\circ}$}
\newcommand{\kms}{km s$^{-1}$}
\usepackage{lscape,graphicx,afterpage}
\begin{document}

\title{The Effect of Cluster Environment on Galaxy Evolution in the
  Pegasus I Cluster.} 

\author{Lorenza Levy\altaffilmark{1} \& James A. Rose}
\altaffiltext{1}{NASA/Jenkins Predoctoral Fellow.}
\affil{University of North Carolina at Chapel Hill}
\affil{Department of Physics and Astronomy, CB 3255, Chapel Hill, NC
  27599}
\email{llevy@physics.unc.edu; jim@physics.unc.edu}
\and
\author{Jacqueline H. van Gorkom} 
\affil{Columbia University}
\affil{Department of Astronomy, 550 W. 120th Street, New York, N.Y. 10027}
\email{jvangork@astro.columbia.edu}
\and
\author{Brian Chaboyer}
\affil{Dartmouth College}
\affil{Department of Physics and Astronomy, 6127 Wilder Laboratory,
  Hanover, NH 03755}
\email{brian.chaboyer@dartmouth.edu}

\begin{abstract}
We present neutral hydrogen observations of 54 galaxies in the Pegasus
cluster. The
observations include single dish HI measurements,
obtained with the Arecibo telescope for all 54 galaxies in the sample, 
as well as HI images, obtained with the VLA for 10
of these. The Arecibo profiles reveal an overall HI deficiency in the
cluster, with $\sim$40$\%$ of the galaxies in the core of the cluster
showing modest deficiencies of typically a factor of 2 -- 3. The HI 
morphology of some galaxies shows that the HI disk is smaller than the
optical disk and slightly offset from the stars.  We find a
correlation between HI deficiency and the ratio of the HI disk size to
optical disk size. More HI deficient galaxies have relatively smaller
HI disks,  a configuration that is
usually attributed to an interaction between the interstellar medium
(ISM) of the 
galaxy and the hot intracluster medium (ICM).  Such a result is
surprising since the
Pegasus cluster has a low level of X-ray emission,
and a low velocity dispersion. The low velocity dispersion,
coupled with the lack of a dense hot ICM indicate that ram pressure
stripping should not play a significant role in this
environment.
In addition, two 
of the galaxies, NGC7604 and NGC7648, are morphologically
peculiar. Their peculiarities indicate contradictory scenarios of what 
is triggering their unusual star formation. H$\alpha$ imaging, along
with long-slit spectroscopy of NGC7648 reveal morphological features
which point to a recent tidal interaction. On the other hand, H$\alpha$ 
imaging of NGC7604 reveals a strong episode of star formation
concentrated into an asymmetric arc, preferentially located on one
side of the galaxy. VLA HI mapping shows the HI also highly
concentrated into that region, suggestive of a ram pressure event. Our 
data indicate that ISM-ICM interactions may play a
role in a wider variety of environments than suggested by simple ram
pressure arguments.
\end{abstract}

\keywords{galaxies: clusters: general, galaxies: evolution, galaxies:
  ISM, radio lines: galaxies}

\section{Introduction}

Understanding the rapid evolution of the star formation
rate in rich clusters of galaxies since z=0.5, first documented by
\citet{bo78,bo84}, remains 
a central issue in  extragalactic  astronomy.  Subsequent
spectroscopy  and HST imaging has revealed a higher  fraction of
spiral  galaxies in distant clusters   
than in present epoch  clusters \citep{dg83,d99}. The heart of the
problem,  then, is to explain the rapid
evolution in the spiral population  since  z=0.5.
It has long  been evident  that the
cluster  environment  is capable of  removing  the 
gas from a galaxy via hydrodynamic interaction between the interstellar
medium (ISM) of cluster galaxies and the pervasive hot intracluster medium
(ICM) \citep{gg72, n82, ss02}.
However,
several tidal perturbation  scenarios have also been suggested that could
drastically  deplete the ISM in spiral  galaxies by inducing large
episodes of star formation \citep{m96,b99}.

It has been generally proposed that hydrodynamic ISM-ICM interaction, 
most specifically the ram pressure, momentum transfer process introduced by
\citet{gg72}, will be most effective in the high velocity-dispersion high
ICM density environments of the central regions of rich clusters.  In
contrast, 
the lower velocity dispersion environment of poor clusters and galaxy groups
has been seen as more conducive to tidal interaction induced effects.
However, subsequent work has revealed a potentially more complex situation.
First, it has been proposed that the cumulative effect of many high speed
tidal encounters in rich clusters can serve to destabilize disk galaxies and
lead to gas removal in the ``galaxy harassment'' process \citep{m96}.  Thus
the tidal encounter hypothesis might play an important role even in rich
clusters.  Second, the role of ISM-ICM interaction may be surprisingly
ubiquitous in lower ICM density and lower velocity dispersion
environments.  For 
example, spiral galaxies with substantial HI depletion have been observed at
such remote distances in the outskirts of the Virgo cluster that it appears
impossible for these galaxies to have passed through the center
of Virgo \citep{s02}, although a detailed analysis of the errors
in distances and depletion factors does raise significant questions about the
status of these galaxies \citep{s04}. 
Third, galaxy starvation, as
proposed by \citet{l80}, removes the HI from a
galaxy when it first enters the cluster, reducing significantly the
star formation lifetimes, and making it more susceptible to ram
pressure effects. 
In addition,
HI imaging of spirals in the low-density Ursa Major cluster reveals large
HI filaments in the vicinity of several spirals, possibly
indicating substantial loss of HI in progress \citep{v04}.  These
results have led 
to the idea of substantial ``preprocessing'' of the ISM of spirals in lower
density environments, although whether the preprocessing is caused by ram
pressure or by tidal interaction remains unclear \citep{v04,vg03}.
In addition, several examples have been found of
spirals in groups or poor clusters which appear to be undergoing HI depletion
and asymmetric star formation that is characteristic of that predicted by the
ram pressure process (e.g., \citet{m93}).  In particular,
the arc-shaped rim of star formation seen in NGC~2276 in the NGC~2300
group, as 
well as the swept back appearance of its radio emission,
\citep{m93} looks similar to the striking examples of spirals in
rich clusters whose optical and radio anomalies are convincingly ascribed to 
ram pressure effects \citep{g95,k04,ovg05}.
Thus there is reason to 
suspect that something other than the classic ram pressure
stripping argument of \citet{gg72} may be important in lower velocity
dispersion environments. Specifically, \citet{gg72} imply that
at a given galactic radius, if the ram pressure exceeds the local restoring
force, then the HI gas is completely stripped at that radius. It may
be the case 
that in lower velocity dispersion environments, some of the HI gas (perhaps
lower density clouds with larger cross-section) is
stripped at a given radius, even if the ram pressure does not exceed
the local restoring force.

In this paper we focus our attention on the nearby Pegasus I cluster of
galaxies. As is further discussed in \S 4,
the low velocity dispersion in the spiral-rich Pegasus I cluster,
coupled with  
the lack of a dense hot ICM, indicates that  ram  pressure  stripping
should  not  play a  significant  role in  this 
environment.  Thus, Pegasus I in principle provides an ideal case
in which to isolate the effects of tidal  perturbations  on the
evolution of galaxies.
 
However, it has recently been shown, contrary to previous studies
\citep{ssb81}, that at
least some spiral galaxies in the Pegasus I cluster have a
deficiency in HI content \citep{s01,gh85}. \citet{gh85} find 3 of 17
spiral galaxies, within 1R$_{A}$ of the Pegasus I cluster center, are
deficient by a factor of 2 or more. \citet{s01} find 10 of 25 spiral
galaxies, within 1R$_{A}$ of the Pegasus I cluster center, have
DEF$\geqslant$0.3.
This fact has important  implications in regards to the
evolution of cluster  galaxies; i.e., 
ISM-ICM interactions may play a significant role in galaxy evolution
in a wider variety of environments than expected from the classic 
\citet{gg72} argument that ISM-ICM effects occur only when the gravitational 
restoring force of the stellar disk is exceeded by the ram pressure
momentum transfer. 

In this paper we provide both single-dish and spatially resolved 
observations of the HI disks of spiral galaxies in the Pegasus I cluster, as
well as a limited amount of optical broadband and H$\alpha$ imaging.  In \S2
we present the observational data, while in \S3 the HI properties of the 
Pegasus I galaxies are derived and summarized.  In \S4 we discuss the HI
deficiencies of the Pegasus I galaxies and put our results into the context
of other studies in \S5.

\section{Observations}

\subsection{Sample of Galaxies}
\begin{figure}[htb]
\plotone{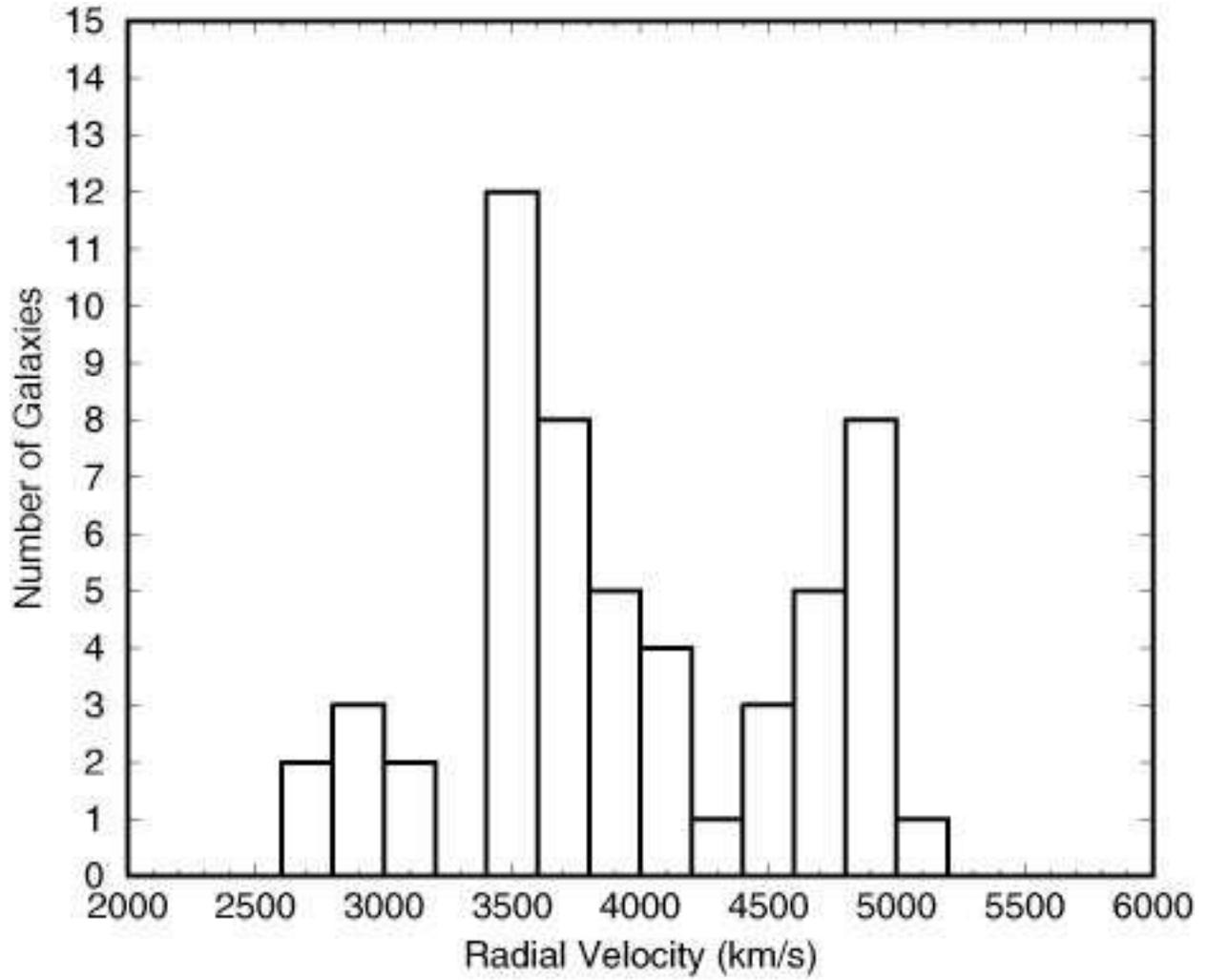}
\figcaption{Velocity histogram. The central group is centered
  at 3900 \kms, the foreground group at 2900 \kms,
  and the background group centered at 5000 \kms.\label{vel}}
\end{figure}

\begin{figure}[htb]
\plotone{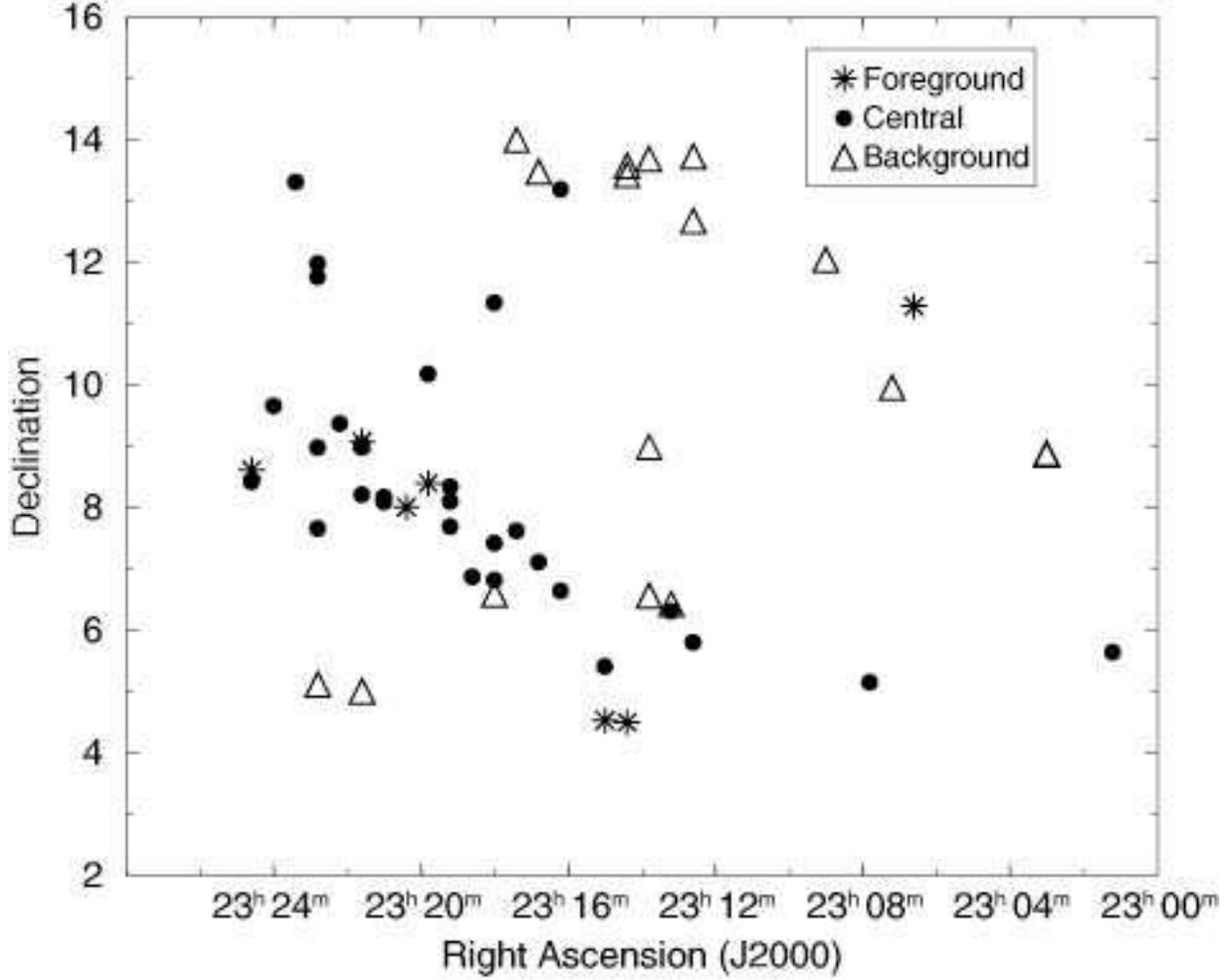}
\figcaption{Spatial distribution of the 54 galaxies in the sample. The
  foreground group is indicated with stars, the central group with
  dots and the background group with triangles.\label{radec}}
\end{figure}

The Pegasus I galaxy cluster was originally delineated by \citet{zkk65}.
The structure of the cluster, and its
separation from the background Pegasus II cluster, was further established
by the optical redshift survey of \citet{cr76}.  Subsequent
studies of the HI content of spirals in Pegasus I
\citep{ssb81,rh82,gh85,s01} have 
provided further redshift data, as well as contradictory claims
regarding depletion of HI in Pegasus I spirals.
We have chosen our sample of galaxies to lie within an RA range of 
23$^{h}$ and 23$^{h}$ 30$^{m}$   
and DEC between 2\degree and 14\degree, following
~\citet{rh82}.
Figure~\ref{vel} shows a velocity histogram for this sample of 54
galaxies. From here, we can distinguish three separate
groupings. There is a central group, with a mean redshift of 3900
\kms, which is situated between a foreground and a background
group. The foreground group has a mean redshift of 2900 \kms and is
composed of 7 disk galaxies, which have velocities between 2500 \kms
and 3400 \kms. 
The central group has a mean redshift of 3900 \kms and 
is composed of 30 galaxies with velocities between 3400 \kms and 4400
\kms. Of these 30 galaxies, 28 of these are disk galaxies with known
morphological types, and 2 of these (NGC7604 and NGC7648) are
morphologically peculiar. These two peculiar galaxies will be
discussed in detail in Section~\ref{pecs}. It is important to note
that the 2 central ellipticals (NGC7619 and NGC7626) fall in the
velocity range of the central group. 
The background group has a mean redshift of 5000 \kms, and is 
composed of 17 disk galaxies with a velocities between 4400 \kms and
6000 \kms. Figure~\ref{radec} shows the spatial distribution of these
galaxies, with the foreground group
indicated with stars, the central group indicated with dots and the
background group indicated with triangles. 
Upon inspection of Figure~\ref{vel} and Figure~\ref{radec}, we see
that the foreground group and the central group are close in
projection on the sky,
while the background group appears spatially displaced
from the other two groupings. It may be the case, then, that the
foreground group is part of the central group, constituting its low
velocity members. The background group may be associated with the main
ridge of the Pisces-Perseus supercluster which lies at a redshift of
between 4000 and 6500 \kms. \citet{hg86} show that the southern region
of the Pisces-Perseus supercluster ridge (DEC$\leqslant$ 35\degree) is
composed of filaments which connect the supercluster with the Local
Supercluster, and that Pegasus is embedded in this narrow filament.
It is also worth noting that, as pointed out by \citet{cr76}, the
Pegasus I galaxies appear flattened on the sky into 
a linear configuration, with many galaxies trailing off to the Southwest
of the two central ellipticals.  This tendency is also evident in
Fig.~\ref{radec}.  In fact, all three redshift components (foreground,
central cluster, and background) appear to follow a NE-SW axis, with
the background component displaced to the Northwest.  Note that the three
galaxies in the background component that lie furthest to the Southeast, and
thus do not follow the general linear trend, are also at the high velocity
end of that component, and may form a separate group.

We also obtained 21 cm observations of non-Pegasus
spiral galaxies with the Arecibo\footnote{The Arecibo Observatory is
part of the National Astronomy and Ionosphere Center, which is
operated by Cornell University under a cooperative agreement with
the National Science Foundation.} telescope during the one hour after
Pegasus I had transited past the zenith angle constraint. 
Altogether,
17 spiral galaxies approximately one hour East of
Pegasus were observed, and are used as a comparison sample for the
Pegasus I galaxies.
Most of 
these galaxies are in the Perseus-Pisces supercluster and are
discussed in more detail in Section 4.

\placefigure{vel_hist_new.eps}
\placefigure{ra_dec_new.eps}

\subsection{Arecibo HI Profiles}
HI 21 cm line observations were obtained with the Arecibo
305 meter telescope of the 54 galaxies in the Pegasus I cluster, and
the 17 non-Pegasus spirals. Forty six
of these Pegasus galaxies, as well as the 17 non-Pegasus I galaxies
that form a reference sample (as described in Section 2.1),
were observed in September 2002 using the  
dual circular L-narrow receiver, with the four subcorrelators covering
a frequency range of 25 MHz, with 2048 channels, resulting in a
resolution of 2.6 \kms.
8 galaxies were observed in
October 2004 using the dual linear 
L-wide receiver, with 1024 channels, and a resolution of 5.3
\kms. Both of these sets of observations were made with a 
beam-width of 3 arcmin at 1415 MHz, and
each galaxy in the sample was observed in total 
power mode with 5 minutes spent ON the galaxy and 5 minutes spent OFF
the galaxy. Total ON-source integration times ranged from 5 to 60
minutes.
Each ON/OFF pair was then averaged together, and boxcar smoothed using
5 channel bins. The baseline for each averaged spectrum was fit by a
polynomial and subtracted, and the integrated HI flux was obtained in
Jy \kms. For each averaged spectra, the rms noise was obtained by
integrating the flux in the baseline, in bins of 400 \kms. For the
frequency range of 25MHz, this resulted in approximately 12 bins
across the baseline (not including the detection). The rms scatter was
calculated for these 12 measurements of the baseline, and this is the
error in the flux, which is on the order of 0.1 Jy \kms.

The Arecibo 21 cm observations were made with the upgraded Gregorian
feed 
system, reducing the uncertainty in the 21 cm fluxes significantly
from the previous observations by \citet{ssb81}, \citet{gh85}, and
\citet{s01}. With the 
improved sensitivity of the Gregorian feed system,   
the uncertainties in the HI fluxes have now been reduced to the point 
where other uncertainties, i.e., optical angular diameters, apparent
blue magnitudes and determination of morphological types, now
represent the principal error in diagnosing the HI deficiency.

\subsection{VLA HI Imaging}
HI 21 cm line observations were made in April and May of 2004, using
the CS configuration of the Very Large Array (VLA)\footnote{The
  National Radio Astronomy Observatory is a facility of the National
  Science Foundation operated under cooperative agreement by
  Associated Universities, Inc.} with spacings
ranging from 0.035 to 3.4 km.  The observations were pointed at nine
galaxies individually and one pointing was centered at the center of
the Pegasus cluster.  Typical observing time on source was 7 hrs. A
nearby phase calibrator, 2255+132, was observed every 45 minutes and
0137+331 (3C48) was observed as flux (16.0 Jy) and bandpass
calibrator.  For the observations of all but one of the individual
galaxies the correlator was configured to cover 3.125 MHz with 63
velocity channels, using online Hanning smoothing. The resulting
channel spacing and velocity resolution is about 10 \kms. For the
central pointing and one of the galaxies, we used a 6.25 MHz bandwidth
and 63 channels with no online Hanning smoothing, resulting in a 20
\kms velocity resolution, but a larger velocity coverage.
Instrumental parameters of the observations are summarized in
Table~\ref{vla_obs}. 
We used NRAO's Astronomical Image Processing System (AIPS) to do the
calibration and imaging. Initially data cubes were made without
continuum subtraction and inspected for HI line emission. In several
of the cubes more than one galaxy was detected. We identified channels
without line emission.  The continuum was subtracted in the UV plane
by making a linear fit through the line-free channels.  Image cubes
were made with various weighting schemes. Here we present results
using uniform weight and robust 1, which optimizes sensitivity, while
still producing a gaussian beam. The resulting angular resolutions are
listed in Table~\ref{vla_obs}. The images were CLEANed. The resulting
rms noise in 
the images is typically 0.35 mJy beam$^{-1}$ or about 10$^{19}$
cm$^{-2}$ per 
10 \kms channel. Total HI images were made by smoothing the cubes
spatially and in velocity, in the smoothed cube we then set all pixels
below 1 or 2 sigma to zero and used this as a mask for the full
resolution cube to calculate the moments.

\placetable{vla_obs}

\subsection{Optical Observations\label{obs_optical}}
We also have optical imaging of NGC7604 and NGC7648. The
optical H$\alpha$, B, and I band images were taken with the 2.4 meter
MDM telescope in September 2001, using the MDM 8K x 8K mosaic CCD
camera, with a pixel size of 15 microns. The scale is 0.206
arcsec/pixel and the images have been binned 2x2, so
the final image scale is 0.41 arcsec/pixel. The B-band and I-band
images have 300 second exposure times and the H$\alpha$ images have
600 second exposure times. The H$\alpha$ images were taken through
both on-line and off-line interference filters, and the final 
H$\alpha$-only images were constructed from the difference between
registered and normalized on-line and off-line exposures.

\section{Derivation of Galaxy Properties}
Of the 54 observed galaxies in the Pegasus I cluster, 52 of these are
spiral
galaxies with known morphological type, and 2 are morphologically
peculiar. The Arecibo HI profiles for the 52 disk galaxies can be
found at http://www.physics.unc.edu/$\backsim$llevy/pegasus. Tables~\ref{one},
and~\ref{two}   
present the data for the 54 Pegasus galaxies and the 17 non-Pegasus
spiral galaxies observed.

Table~\ref{one} includes the physical properties of the observed 
galaxies.

\textit{Columns 1 and 2}: galaxy name; (1) NGC or alternate name; (2)
UGC name.  

\textit{Columns 3 and 4}: RA and DEC in J2000.0 coordinates obtained
from the NASA Extragalactic Database (NED)\footnote{NED is
  operated by the Jet Propulsion Laboratory, California Institute of
  Technology, under contract with the National Aeronautics and Space
  Administration.}.

\textit{Column 5}: heliocentric 21 cm velocity, in \kms, obtained from
NED. The NED velocities agree well with the Arecibo profiles, except
for NGC7615, where we use the velocity obtained from the Arecibo
profile. 

\textit{Column 6}: morphological type, in de Vaucouleurs (RC3)
notation, ignoring the presence of bars. Where the superscripts are:
(a) obtained using the RC3 catalogue, (b) measured by the authors
using the Palomar Sky Survey prints.

\textit{Column 7}: apparent blue magnitude corrected for galactic
extinction, internal extinction, and redshift correction as prescribed
in \citet{b96}. 
The superscripts are (a) corrected magnitude obtained
using the RC3 catalogue, the error in magnitude is 0.13 mag
\citep{b96}, (b) the uncorrected magnitude is obtained using the UGC,
and corrected as prescribed in \citet{b96}, (c)
uncorrected magnitude, obtained from the Zwicky catalog, and corrected
as prescribed in \citet{b96}, (d) uncorrected magnitude, obtained
from the Flat Galaxy Catalogue, and corrected as prescribed in
\citet{b96}. 

\textit{Column 8}: blue semi-major axis, in arcminutes. The
superscripts are: (a) obtained using the UGC catalogue, where the
error in the semi-major axis is 15$\%$ \citep{hg84}, (b) measured by
the authors using the Palomar Sky Survey prints.

\textit{Column 9}: ratio of the semi-major axis to the semi-minor
axis: b/a.

\textit{Column 10}: velocity width, in \kms, defined as 20$\%$ of the
peak and determined from the Arecibo HI profiles.

\placetable{one}

The HI deficiencies and other derived quantities are computed for
these galaxies and shown in Table~\ref{two}. 

\textit{Columns 1 and 2}: galaxy name; (1) NGC or alternate name; (2)
UGC name. An asterisk following the name indicates the Arecibo data
was taken on the second observing run in October 2004. 

\textit{Column 3}:  morphological type, taken from column 6 of
Table~\ref{one}. 

\textit{Column 4}: total exposure time, in minutes, for the Arecibo HI
profiles. 

\textit{Column 5}: 21 cm flux, in Jy \kms, obtained from the
Arecibo HI profiles, corrected for
pointing errors and aperture adjustments. The error in the pointing is 
less than 5 arcseconds. The aperture correction is 2$\%$. 

\textit{Column 6}: log of the HI Mass, where:
\begin{displaymath}
(M_{HI}/M\sun) = 2.36 \times 10^{5} (F/Jy\;km\;s^{-1}) (D/Mpc)^{2}
\end{displaymath}
where the distance to each galaxy is taken as the distance to the
Pegasus I cluster center. Using an H$_{0}$ of 100 \kms
Mpc$^{-1}$, results in a distance to Pegasus I of 40 Mpc.

\textit{Column 7}: log of the blue luminosity, in L$\sun$, with
L$_{B\sun}$=5.37 \citep{sk57}, and with m$_{0}$ from column 7 in
Table~\ref{one}.

\textit{Column 8}: log of the linear optical diameter squared, in
kpc$^{2}$. The linear optical diameter is the UGC blue semi-major axis
(from Table~\ref{one}) converted to kpc \citep{sgh96}:
\begin{displaymath}
(D_{0}/kpc)=0.291(D/Mpc)(a/arcmin)
\end{displaymath}

\textit{Column 9}: HI deficiency factor calculated following the
method prescribed by \citet{sgh96} (hereafter SGH96).
\begin{displaymath}
DEF=log(M_{HI})_{exp}-log(M_{HI})_{obs}
\end{displaymath}
where the expected value is calculated using Table 2 in SGH96, and a
DEF greater than zero is HI deficient. 

\placetable{two}

\section{HI Deficiency}
The primary goal of this paper is to ascertain whether spiral
galaxies in the Pegasus I galaxy cluster exhibit deficiencies in their
HI content, as is so evident in the case of spirals in richer clusters
with a well-developed hot ICM (e.g., Coma \citep{ga87} and Virgo
\citep{ca94}). As was mentioned in \S 1, it is generally
considered that HI deficiencies in cluster 
spirals are primarily caused by ram pressure stripping of the galaxy's ISM as
it impacts the hot ICM, as originally proposed by \citet{gg72}.
The case for ram pressure stripping is made when the ram pressure,
characterized
by $\rho_{ICM}v^{2}$ (where $\rho_{ICM}$ is the density of the ICM
and $v$ is the
typical velocity of a galaxy in the cluster), exceeds the gravitational
restoring force of the disk.  In Table~\ref{groups} we summarize the
relevant data for
several clusters of different richness, to place the Pegasus I cluster into
perspective.  In column (2) we list the percentages of early-type vs late-type
galaxies.  In columns (3), (4), and (5) are given the line-of-sight velocity
dispersion in the cluster, the cluster X-ray luminosity, and the central
electron density, inferred from fitting to the X-ray data.  In column (6),
the ram pressure is given.  For reference, the gravitational restoring force
in a spiral disk typically amounts to $\sim$1000 (\kms)$^{2}$ cm$^{-2}$.
Thus while ram pressure stripping is expected to be effective in the
{\it central} regions of Coma and Virgo, it should fail by at least a factor
of 50 in Pegasus I.  Consequently, we do not, a priori, expect to see stripped
spirals in Pegasus I.

\placetable{groups}

\begin{figure}[htb]
\plotone{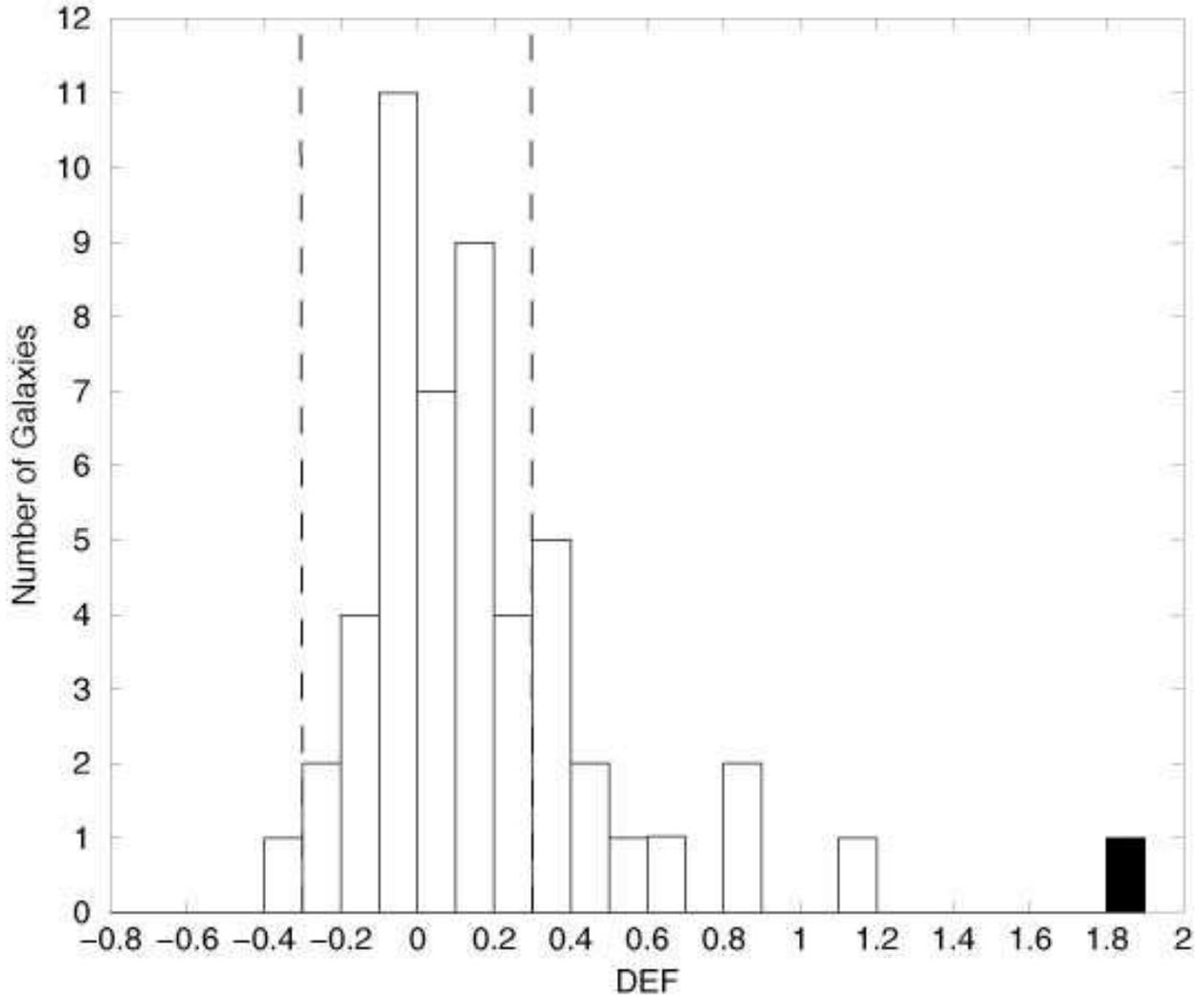}
\figcaption{DEF for the 51 spiral galaxies in the sample. The filled
  rectangle represents NGC7563 which is a non-detection, and its value 
  of DEF is a lower-limit. The vertical
  dashed lines are placed at a factor of two deficiency (DEF=0.3) and
  surplus (DEF=-0.3).\label{hg96_all}} 
\end{figure}

Before discussing HI deficiencies in Pegasus I spirals, we require a
clear idea as to what constitutes an HI deficient galaxy.
In what follows, we use the definition of HI deficiency
specified in \citet{sgh96}, while considering the effect of using other
deficiency definitions in \S4.1.
With that in mind we have examined the histogram of HI deficiencies for
all of the Pegasus I spirals observed by us at Arecibo.\footnote{We have
eliminated the Seyfert 1 galaxy  NGC7469 from further consideration, since
this galaxy shows strong HI self-absorption in its Arecibo profile,
resulting in a spuriously high HI deficiency factor.  Elimination of
NGC7469 leaves us with 51 spirals in our sample.}  That histogram
is plotted in Figure~\ref{hg96_all}. The filled rectangle represents
NGC7563 which is a non-detection, and its value of DEF is a lower-limit.
It is evident from Figure~\ref{hg96_all}
that there is a substantial ``cosmic'' scatter in HI content, but that there 
is also an asymmetric tail to positive HI deficiencies, thus indicating
that at least some spirals in Pegasus I are HI deficient.  Given the rather
continuous nature of the tail to positive DEF factors, however, it is not 
clear from the outset if there is a particular DEF value above which one can
make a strong case for HI deficiency, and if so, what that DEF value is.
Thus in what follows, we first argue that there is indeed a clear case of HI
deficient spirals in Pegasus I, particularly in the central region of the
cluster.  Then we make a statistical case for a deficiency factor of DEF=0.3
as the dividing line for spirals which are highly likely to be HI deficient.

\begin{figure}[htb]
\plotone{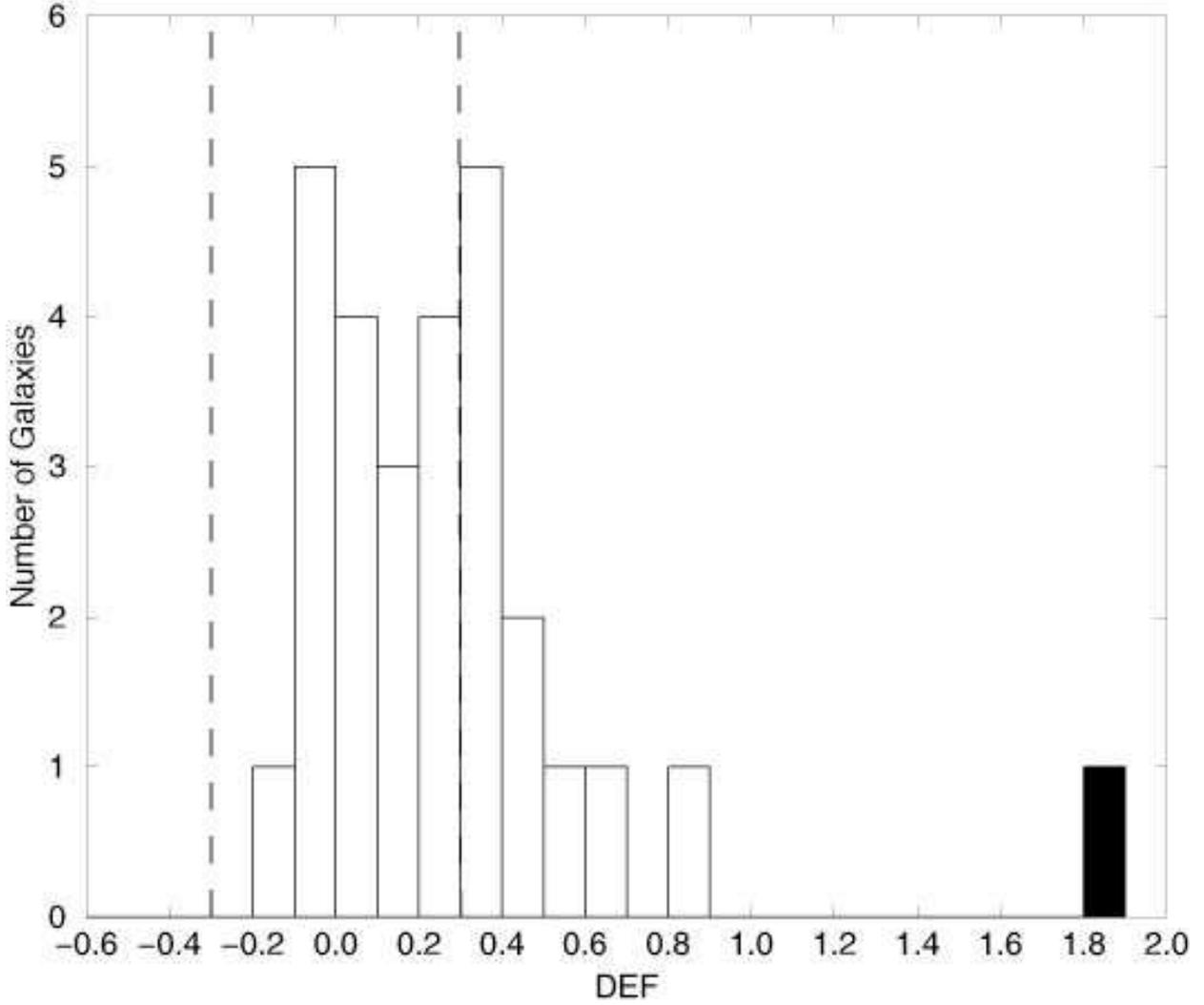}
\figcaption{DEF for the 28 disk galaxies in the central group. The
  filled rectangle represents NGC7563 which is a non-detection, and
  its value of DEF is a lower-limit. The vertical
  dashed lines are placed at a factor of two deficiency (DEF=0.3) and 
  surplus (DEF=-0.3).\label{hg96_core}} 
\end{figure}

\begin{figure}[htb]
\plotone{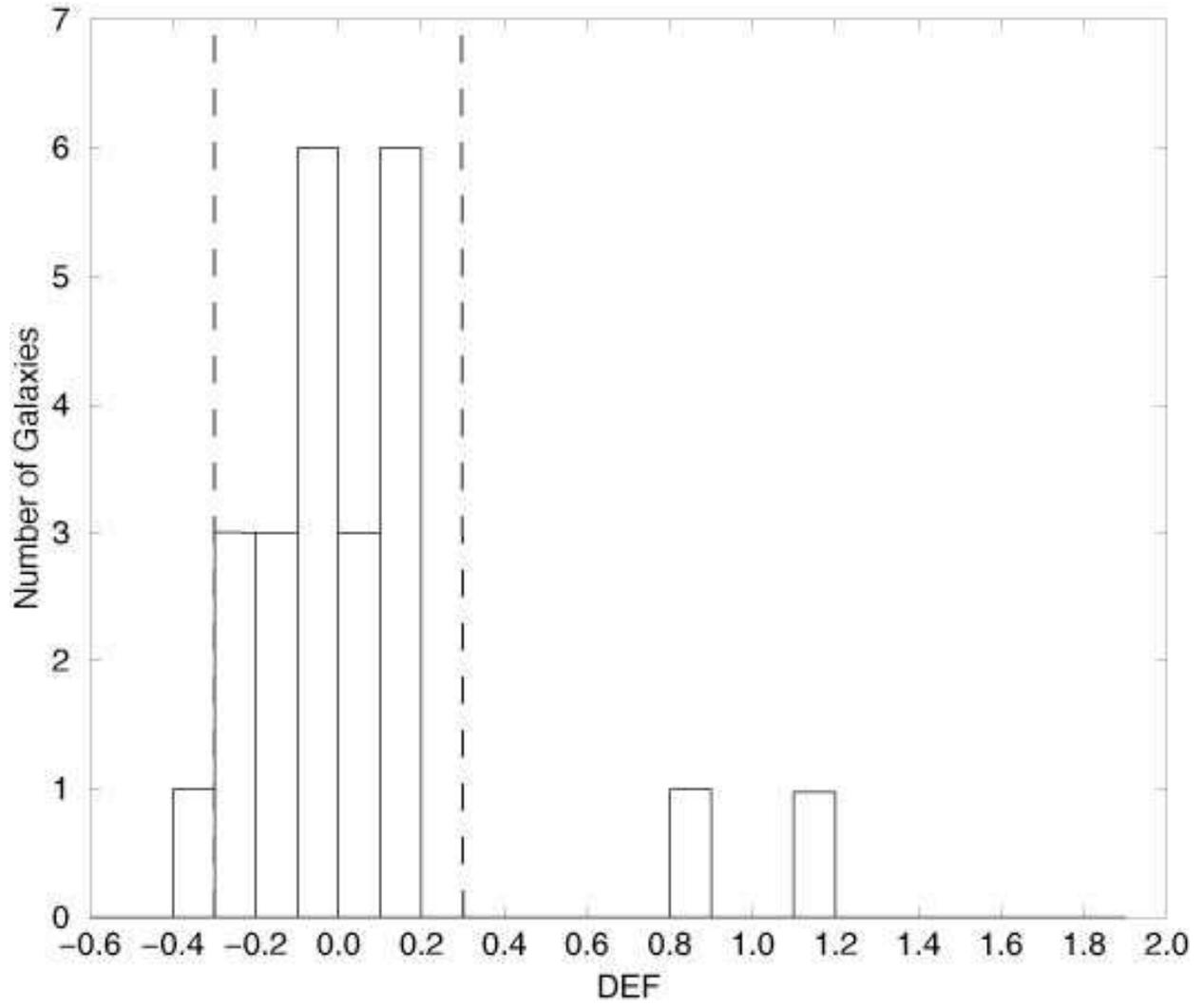}
\figcaption{DEF for the 23 galaxies in the foreground and
  background groups. The vertical
  dashed lines are placed at a factor of two deficiency (DEF=0.3) and
  surplus (DEF=-0.3).\label{hg96_noncore}} 
\end{figure}

To begin with, we make a comparison between the DEF factors for spirals in the
central region of Pegasus I versus those in the foreground and background
groups.  In Figure~5 is plotted the histogram of deficiency factors for the 28
disk galaxies in the central Pegasus I cluster.  The filled rectangle
represents 
NGC7653, which is a non-detection, thus its DEF value is a lower-limit.  The
offset in the distribution of DEF values towards positive deficiency factors
is more pronounced than for the whole sample plotted in Figure~4.  Of the
28 disk galaxies in the central cluster, 6 have DEF$\gtrsim$0.4 and 6 have
0.3$\leqslant$DEF$\leqslant$0.4.  In contrast, the histogram of DEF
factors for 
the 23 spirals in the foreground and background groups is plotted in
Figure~\ref{hg96_noncore}. 
In this case the evidence for HI deficiency is substantially weaker,
particularly when compared with the that for the central cluster shown in
Figure~\ref{hg96_core}.  Of the 23 spirals in the foreground and
background groups, 
only two, 
i.e., 9\% of the sample, have DEF$\geqslant$0.3, as opposed to 40\% of the
central cluster sample (12 out of 28).  In addition, the non-parametric 
Kolmogorov-Smirnov two-sample test, when applied to the DEF factors in the 
central group versus those in the foreground and background group, rejects
the hypothesis that the two samples are
drawn from the same parent population at the 97\% confidence level.
In short, the spirals in the central cluster on average have greater
DEF factors 
than their counterparts in the foreground and background groups.

We can make a second control test for HI deficiency in the central cluster
galaxies by comparing with the DEF factors for the sample of 17
spirals that we 
observed approximately 1 hour East of Pegasus I, which were accessible to us
after Pegasus I had transited through the observing window for Arecibo, and
before our observing sessions had ended.  Centered at RA$\sim
0^h20^m$, many of 
these spirals are located within the Pisces-Perseus Supercluster
(PPS), as can  
be seen in Figure~1 of \citet{hg86}.  In fact, of the
17 spirals, 
11 of them are within the spatial and velocity limits of the PPS.  And
of those 
11 spirals, 9 of them lie within the spatial and velocity limits of a the
cluster 0019+2207 \citep{hm92}.  The basic information on the
HI content 
and DEF factors of these 17 galaxies is compiled in Tables~2 and
3.  The results are 
as follows.  There is a single galaxy at low redshift (cz$\sim$2300
\kms) which 
has DEF=-0.07.  There is a group of 5 galaxies clustered at a mean redshift
of cz$\sim$4500 \kms, which have a median DEF=0.09, and which are foreground
to the PPS.  There are two galaxies with cz$>$5000 \kms that are
likely located 
in the PPS, but not in the cluster 0019+2207, which have DEF=0.13 and
DEF=0.46. 
Finally, the 11 spirals in 0019+2207 have median DEF=0.48.  While we are
dealing with small numbers of galaxies, the results indicate that for galaxies
outside the PPS, there is no evidence for significant Hi depletion, while for
the galaxies situated in the the cluster 0019+2207 within the PPS, there is
evidence for HI deficiency.  Furthermore, 0019+2207, like Pegasus I,
is not a rich cluster.  If one takes the velocity data
for the 10 galaxies given in \citet{hm92}, the mean cluster redshift is
$cz=5832$ \kms, with a 1 $\sigma$ dispersion of $\pm$336 \kms, while if
one galaxy with high velocity is rejected, the remaining 9 galaxies
high a mean 
redshift $cz=5745$ \kms, with a 1 $\sigma$ dispersion of $\pm$203 \kms.
Hence 0019+2207 has a low velocity dispersion, comparable to that of the core
of Pegasus I, and yet also exhibits evidence for HI depletion.  

While the above analysis indicates that Pegasus I spirals in the cluster core
have a skewed distribution to positive DEF values, it is somewhat problematic
to determine what constitutes a significant HI depletion level,
especially since 
there is a certain degree of both observational error and cosmic
scatter in the 
DEF data.  To separate
real HI depletions from cosmic scatter, we have analyzed the distribution
of DEF factors following three approaches.  In all cases a gaussian was
fit to the distribution of DEF factors. First, a gaussian fit with
a $\pm$2.5-sigma clipping resulted in an rms 
scatter in the deficiency of $\pm$0.26 (with a mean of DEF=0.13)
for the full sample of 51 foreground, 
central, and background groups, $\pm$0.17 (mean DEF=0.15) for the
central group  
alone, and $\pm$0.15 (mean of -0.03) for the sample containing the foreground 
and background groups.
If, instead, a 2-sigma clipping is applied,
an rms scatter in the deficiency of $\pm$0.15 for the
foreground, central, 
and background groups as a whole, $\pm$0.12 for the central group, and
$\pm$0.13 for the foreground and background groups sample is obtained.
Finally, since it is evident the above discussion that the distribution
of DEF values is skewed on the positive side by galaxies with HI depletion,
a gaussian was fit to only the negative DEF values, which have a well behaved
distribution, and thus are likely to give a better estimate of the cosmic
scatter in DEF values. This one-sided fit resulted in an rms
scatter in the deficiency of $\pm$0.14 for the complete sample of foreground, 
central, and background groups, $\pm$0.04 for the central group, and
$\pm$0.12 for the foreground and background groups sample.  Based on the above
information, we consider that the 1 $\sigma$ cosmic scatter in our DEF values 
is no larger than $\pm$0.15, and assume this value for the scatter.  
Consequently, any galaxy with a DEF value of 0.3 or greater has a high 
probability of being a truly HI-deficient galaxy, given that it is greater
than 2 $\sigma$ from the mean.  For the remainder of the paper, we will 
consider a deficiency factor DEF$\gtrsim$0.3 (i.e., an HI depletion of
a factor 
of two or greater) to represent the line at which we can confidently
argue for  
a real HI deficiency.  Naturally, some of the galaxies with slightly lower
deficiencies than DEF=0.3 are likely also to be HI-deficient.
However, in those 
cases one cannot make a convincing case for HI deficiency in any one specific
galaxy.

Finally, it is important to note that in Figure~\ref{hg96_all}
the distribution
is centered around zero, but with an asymmetric tail towards positive
deficiency values. In Figure~\ref{hg96_core}, the distribution is
centered around 0.1 and also shows an asymmetric tail towards positive
deficiency values.
In richer, higher-$\sigma$ clusters that are
believed to have undergone extensive ram
pressure stripping, such as Virgo, the mean HI deficiency is
centered around larger positive values (DEF$\sim$0.4), and, in the
case of Virgo, around
75$\%$ of the spiral galaxies in the cluster are HI deficient
\citep{s01}.  Thus there is a clear difference between the typical
level of HI  
depletion found for Pegasus I spirals and that found for spirals in richer
clusters.

\placefigure{hist_96_all-seyfert_new.eps}
\placefigure{hist_96_core_new.eps}
\placefigure{hist_96_noncore-seyfert_new.eps}

\begin{figure}[htb]
\plotone{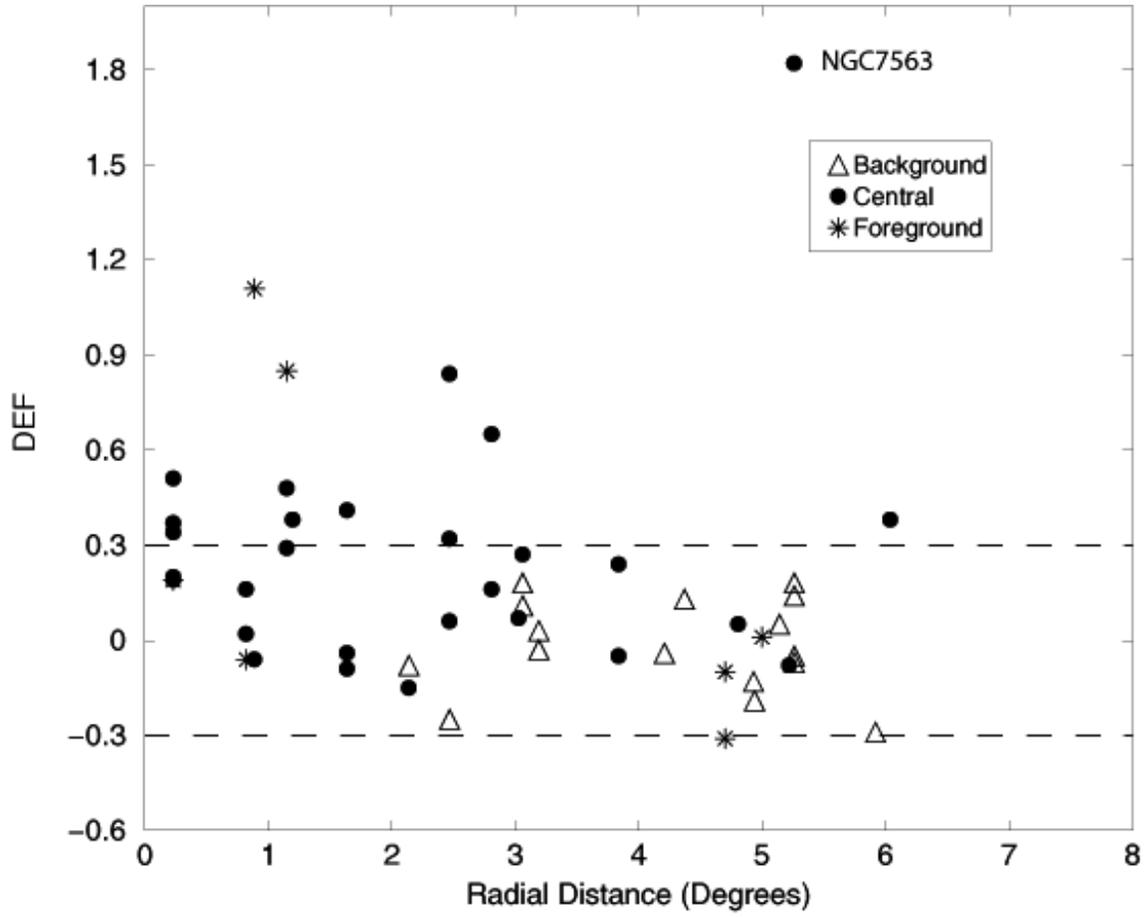}
\figcaption{Distribution of DEF with radial distance from the cluster
  center, as determined by the two central ellipticals.\label{rdef}}
\end{figure}

A plot of how deficiency factor varies with radial distance from the
center of the cluster is shown in
Figure~\ref{rdef}. The
galaxies in the foreground group are indicated with stars, the central
group with dots and the background group with triangles. The
horizontal dashed lines mark where DEF=$\pm$0.30.
 We see that most of the galaxies whose deficiencies are greater
than a factor of two are located within the inner 2.5\degree. The two
outlying
highly deficient galaxies at around 5\degree are the central group
galaxy NGC7563, and the background group galaxy NGC7469. NGC7563 is an
Sa galaxy, with a companion UGC12463 at 3.7 arcmin. From the HI
profile, we see that no HI is detected, and thus the DEF 
value is a lower limit.

\begin{figure}[htb]
\plotone{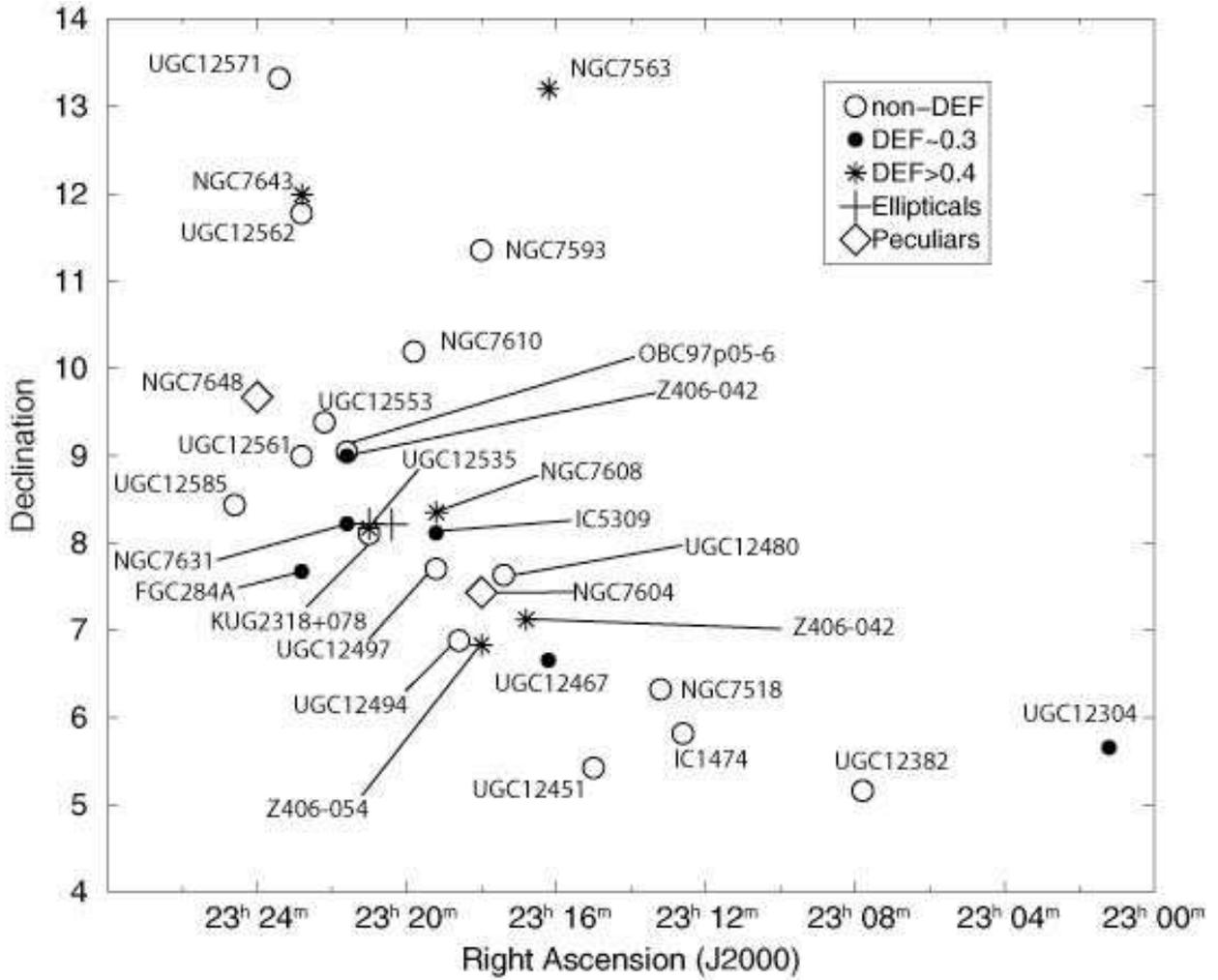}
\figcaption{Spatial distribution of the galaxies in the central Pegasus
  I cluster.\label{radec_core}}  
\end{figure}

Figure~\ref{radec_core} shows the spatial distribution of the galaxies
in the central group of
our sample. The two central ellipticals, NGC7619 and NGC7626 are
indicated with crosses. The two morphologically peculiar galaxies,
which will be discussed in detail in \S6, are indicated
with diamonds.
The non-deficient galaxies are marked with open circles, the slightly
deficient galaxies are marked with filled circles, and the highly
deficient galaxies are indicated with stars.

\subsection{Comparison with Previous Results}\label{sect:previous}
Previous studies of Pegasus I spirals \citep{ssb81,gh85,s01} have produced
somewhat contradictory results about HI deficiencies, with earlier results
tending to find no HI deficiencies while \citet{s01} do indeed find
significant HI depletions.  Part of this change in perspective is due to
observations of increased accuracy.  Since most of the observed deficiencies
in Pegasus turn out to be at a modest (factor of 2) level, high quality
data is required to make such depletions evident. 
In addition, the sample of galaxies used to define the Pegasus I
cluster has differed, producing somewhat different results for the
overall HI deficiency of the cluster. \citet{ssb81} and \citet{gh85}
used a sample of galaxies that extends far out from the cluster
center, possibly including galaxies not associated with the
cluster. They both find no evidence for HI deficiency. \citet{s01}
calculates the HI deficiency for a large, extended sample, similar to
\citet{gh85}, and for the Pegasus galaxies located within 1R$_{A}$ of
the cluster core. The extended sample shows no evidence of HI
deficiency, but the sample within 1R$_{A}$ of the core shows about
40\% of the galaxies to be HI deficient. 
Our results agree very closely with those obtained by 
\citet{s01}, i.e., that in the central core of the Pegasus I cluster, 
$\sim$40$\%$ of the spirals exhibit modest HI depletions. In the
foreground and 
background groups, the incidence of deficiency is considerably lower, which
also helps to explain the fact that earlier results \citep{ssb81,gh85} did 
not find evidence for HI depletion, since those studies were not concentrated
on the Pegasus I core.

\placefigure{def_r_96-seyfert_new.eps}
\placefigure{ra_dec_core_new.eps}

\section{HI Imaging}
\begin{figure}
\plotone{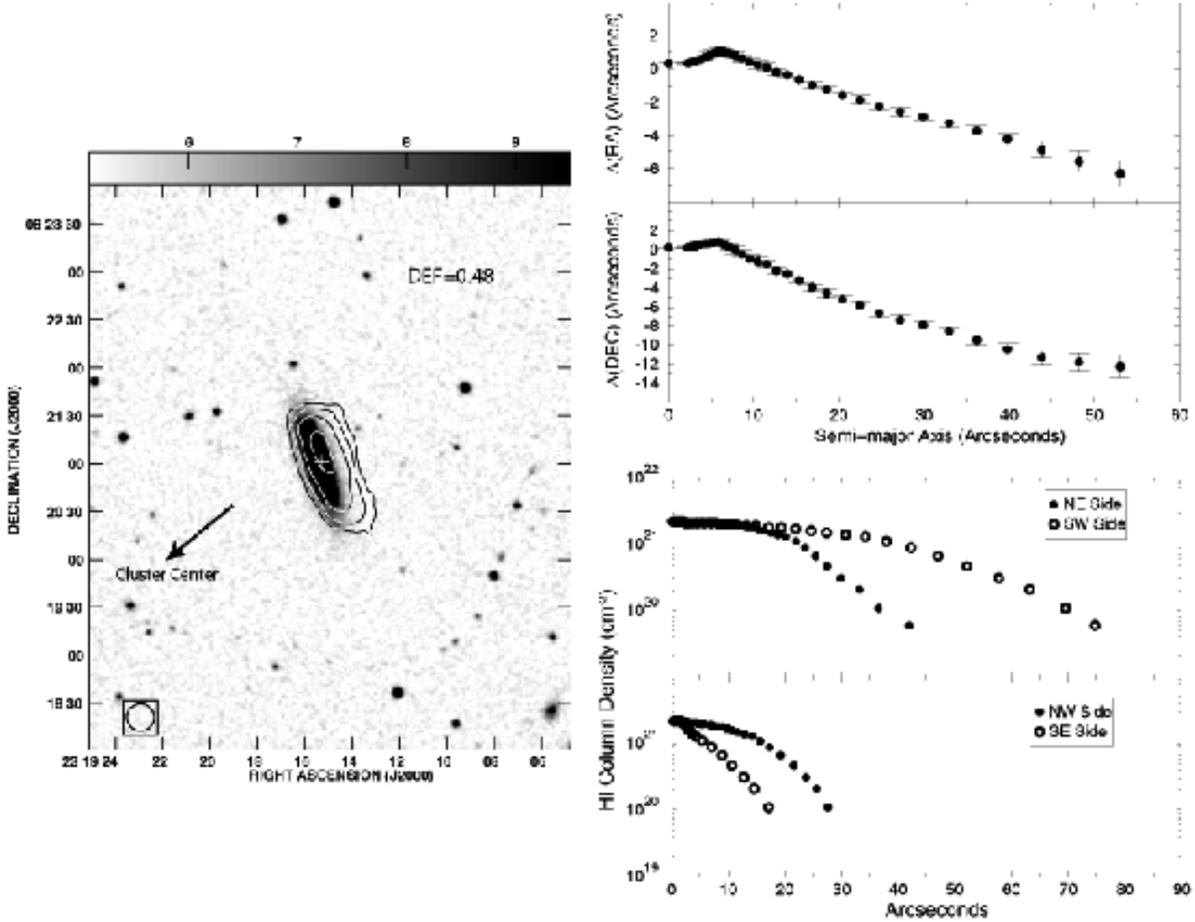}
\caption{\textbf{Left:} HI contours overlaid on DSS for NGC7608. Note
  the HI disk is less extended than the optical disk as well as the
  displaced HI gas and asymmetric contours. 
  The lowest contour is 50
  mJy beam$^{-1}$ \kms which corresponds to 2.4x10$^{20}$
  cm$^{-2}$. The contour levels are at 2.4, 4.7, 9.4 and
  18.9x10$^{20}$ cm$^{-2}$. The beam size is shown in the bottom left
  hand corner.
 \textbf{Top Right:} The
  position center shifts of the ellipses fitted to the HI column density
  levels in NGC7608, measured in arcseconds, are plotted as a function of
 the semi-major axis. There is a shift towards the West and the
  South. \textbf{Bottom Right:} The radial HI column density
  distribution for NCG7608 along the major axis (top) and the minor
  axis (bottom). For the major axis, the filled circles represent the
  distribution from the galaxy center towards the North-East, and the
  empty circles represent the distribution from the galaxy center
  towards the South-West. For the minor axis, the filled circles
  represent the distribution towards the North-West, and the empty
  circles represent the distribution towards the
  South-East.\label{n7608_comp}} 
\end{figure}

\begin{figure}[htbp]
\plotone{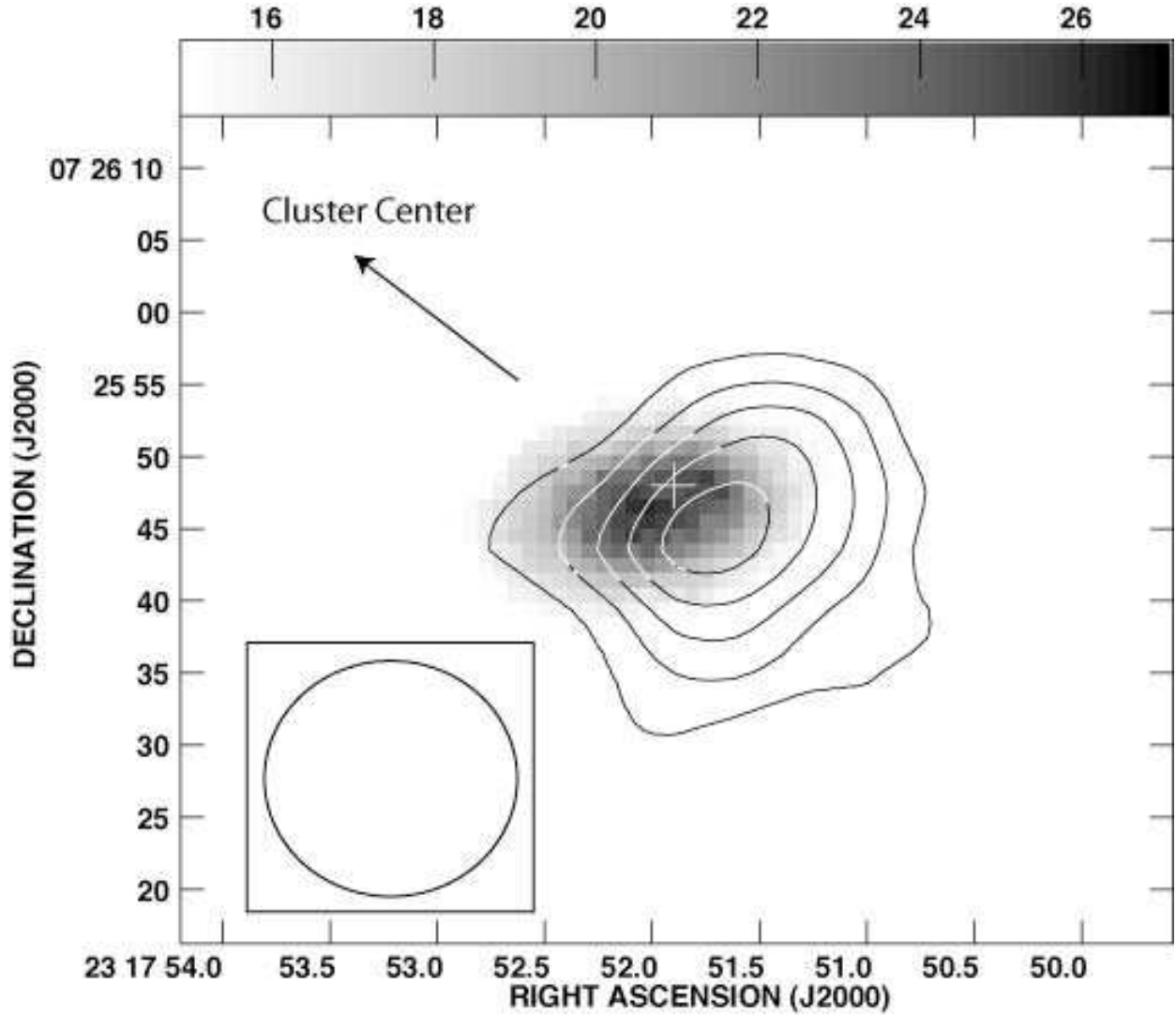}
\figcaption{HI contours overlaid on DSS for NGC7604. Notice the HI
  disk offset from the optical disk.
  The lowest contour is 25
  mJy beam$^{-1}$ \kms which corresponds to 1.2x10$^{20}$
  cm$^{-2}$. The contour levels are at 1.2, 2.4, 3.7, 4.9 and
  6.1x10$^{20}$ cm$^{-2}$.
\label{n7604_vla}}
\end{figure}

\begin{figure}[htbp]
\plotone{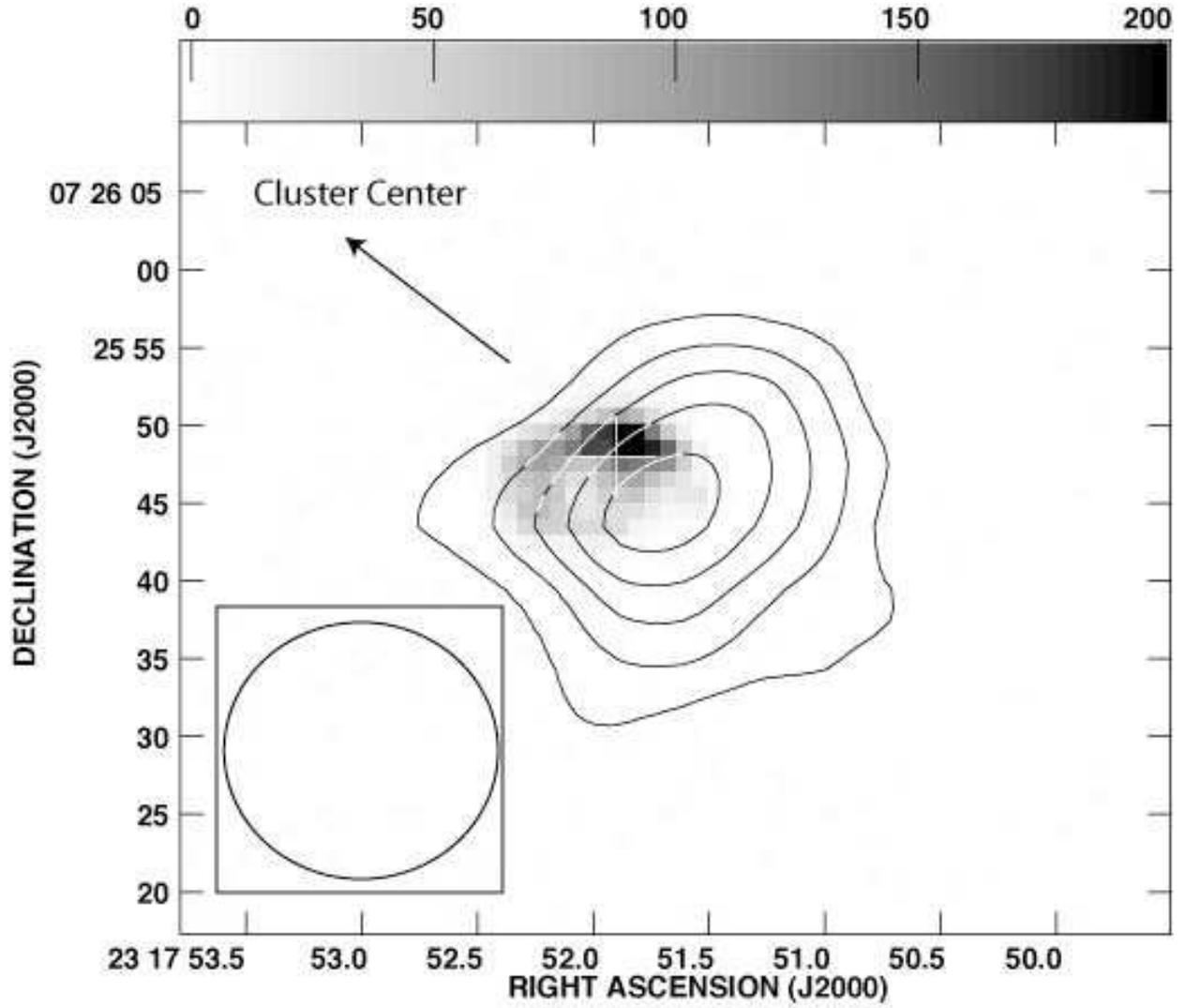}
\figcaption{HI contours overlaid on the H$\alpha$ image for
  NGC7604. The contour levels are the same as for
  Figure~\ref{n7604_vla}. The HI emission is clearly displaced to the
  SW of the arc 
  of H$\alpha$ emission that is concentrated along the NW edge of the
  galaxy.\label{n7604ha_vla}}
\end{figure}

\begin{figure}[htbp]
\plotone{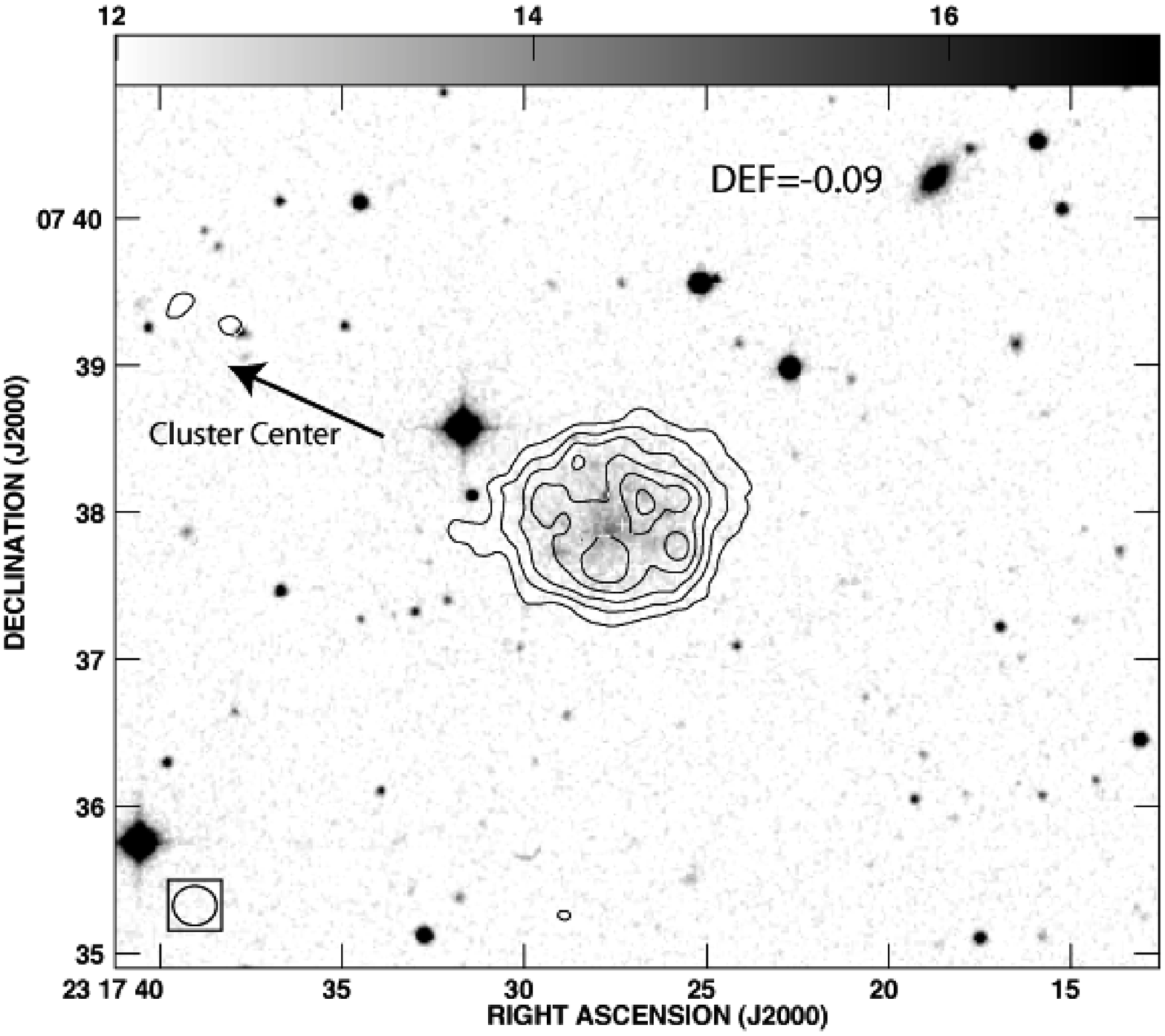}
\figcaption{HI contours overlaid on DSS for UGC12480.
  The lowest contour is 30
  mJy beam$^{-1}$ \kms which corresponds to 1.5x10$^{20}$
  cm$^{-2}$. The contour levels are at 1.5, 2.9, 4.4, 5.8, 7.3 and
  8.8x10$^{20}$ cm$^{-2}$.
\label{u12480_vla}}
\end{figure}

\begin{figure}[htbp]
\plotone{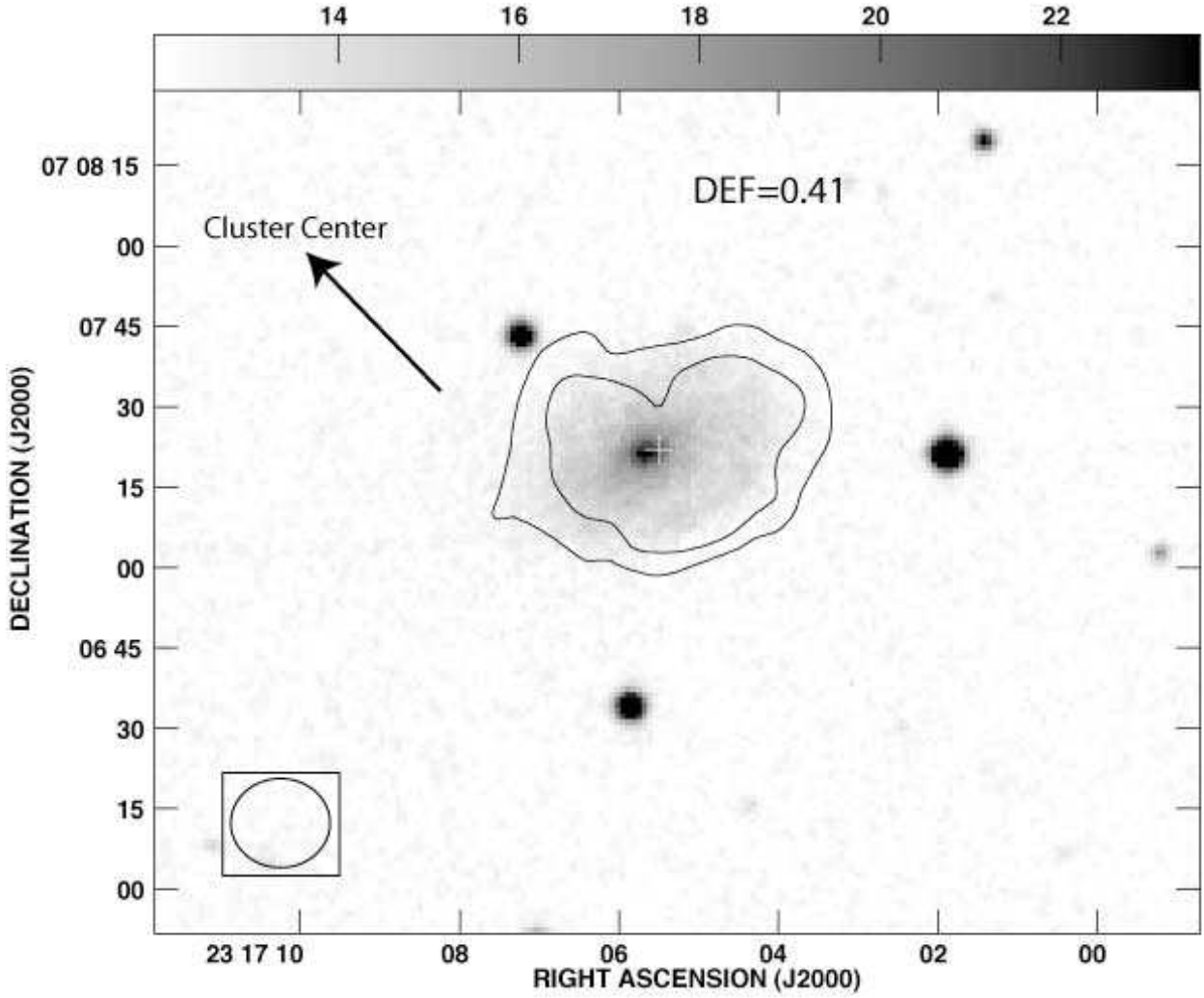}
\figcaption{HI contours overlaid on DSS for Z406-042.
The lowest contour is 50
  mJy beam$^{-1}$ \kms which corresponds to 2.3x10$^{20}$
  cm$^{-2}$. The contour levels are at 2.3, and 
  4.7x10$^{20}$ cm$^{-2}$.
\label{z406_vla}}
\end{figure}

\begin{figure}
\plotone{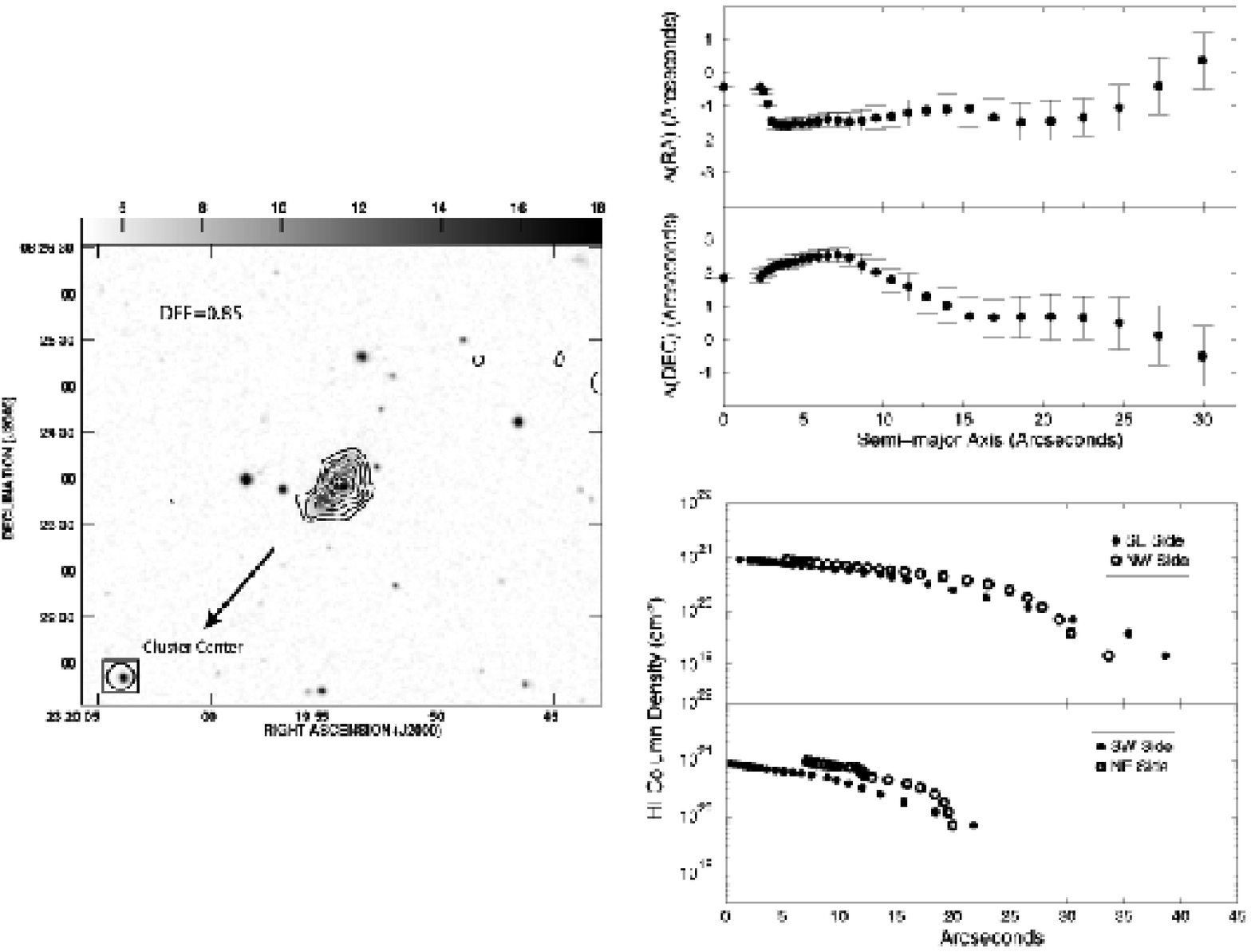}
\caption{\textbf{Left:} HI contours overlaid on DSS for NGC7615. There
  appears to be an asymmetry in the HI contours towards the SE and SW
  sides.
  The lowest contour is 30
  mJy beam$^{-1}$ \kms which corresponds to 1.4x10$^{20}$
  cm$^{-2}$. The contour levels are at 1.4, 2.8, 4.2, 5.6, 7.1 and
  8.4x10$^{20}$ cm$^{-2}$.
  \textbf{Top Right:} The 
  position center shifts of the ellipses fitted to the HI column density
  levels in NGC7615, measured in arcseconds, are plotted as a function of
  the semi-major axis.
  There is a small HI displacement towards the
  West and South, as indicated by the shifts in the ellipse centers.
 \textbf{Bottom Right:} The radial HI column density 
  distribution for NCG7615 along the major axis (top) and the minor
  axis (bottom). For the major axis, the filled circles represent the
  distribution from the galaxy center towards the South-East, and the
  empty circles represent the distribution from the galaxy center
  towards the North-West. For the minor axis, the filled circles
  represent the distribution towards the South-West, and the empty
  circles represent the distribution towards the
  North-East.\label{n7615_comp}} 
\end{figure}

\begin{figure}[htbp]
\plotone{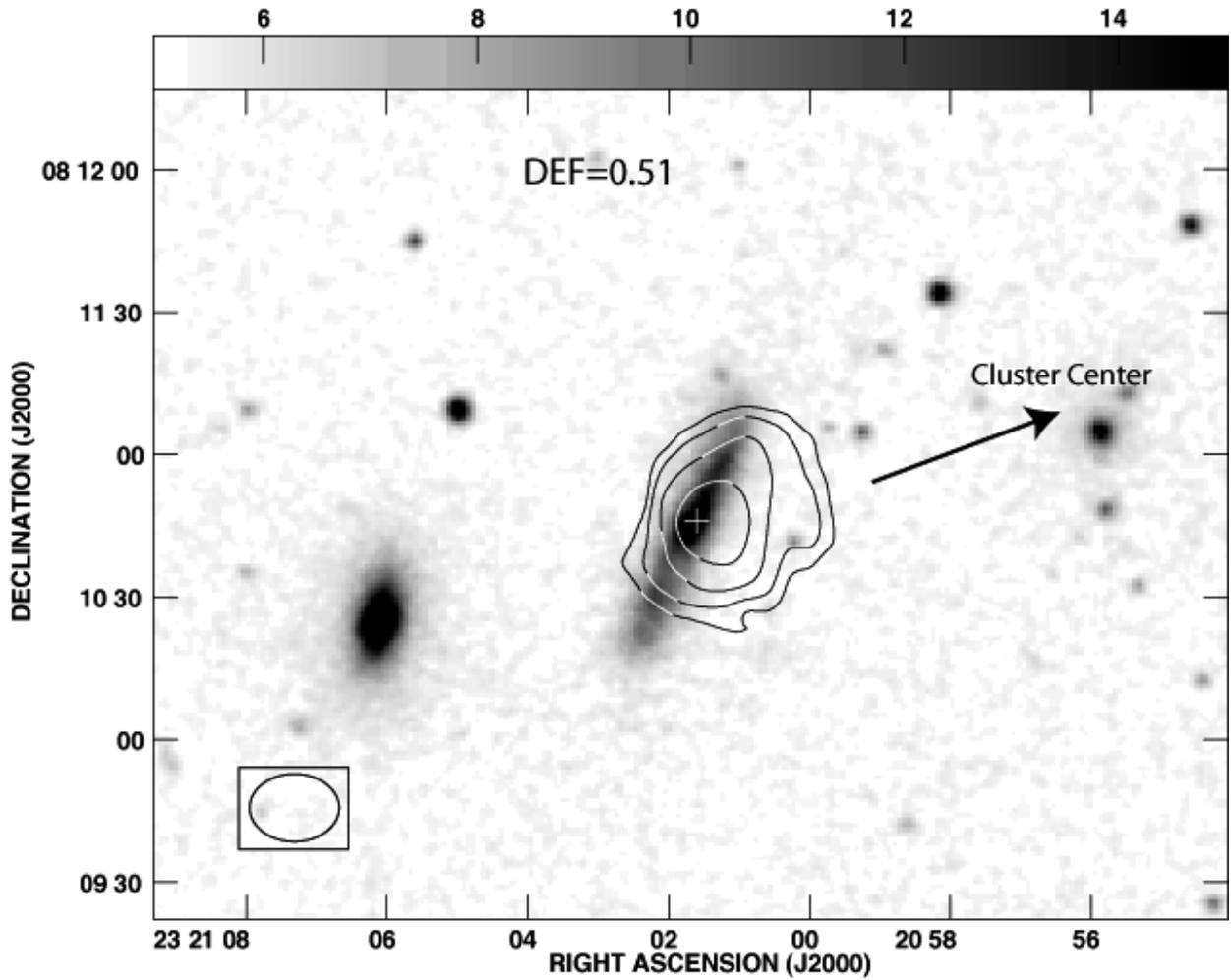}
\figcaption{HI contours overlaid on DSS for UGC12535. The low velocity
  range is completely covered, but the high velocity range is
  missing. The lowest contour is 50
  mJy beam$^{-1}$ \kms which corresponds to 2.6x10$^{20}$
  cm$^{-2}$. The contour levels are at 2.6, 5.3, 11.1, and
  21.2x10$^{20}$ cm$^{-2}$.
\label{u12535_vla}} 
\end{figure}

\begin{figure}
\plotone{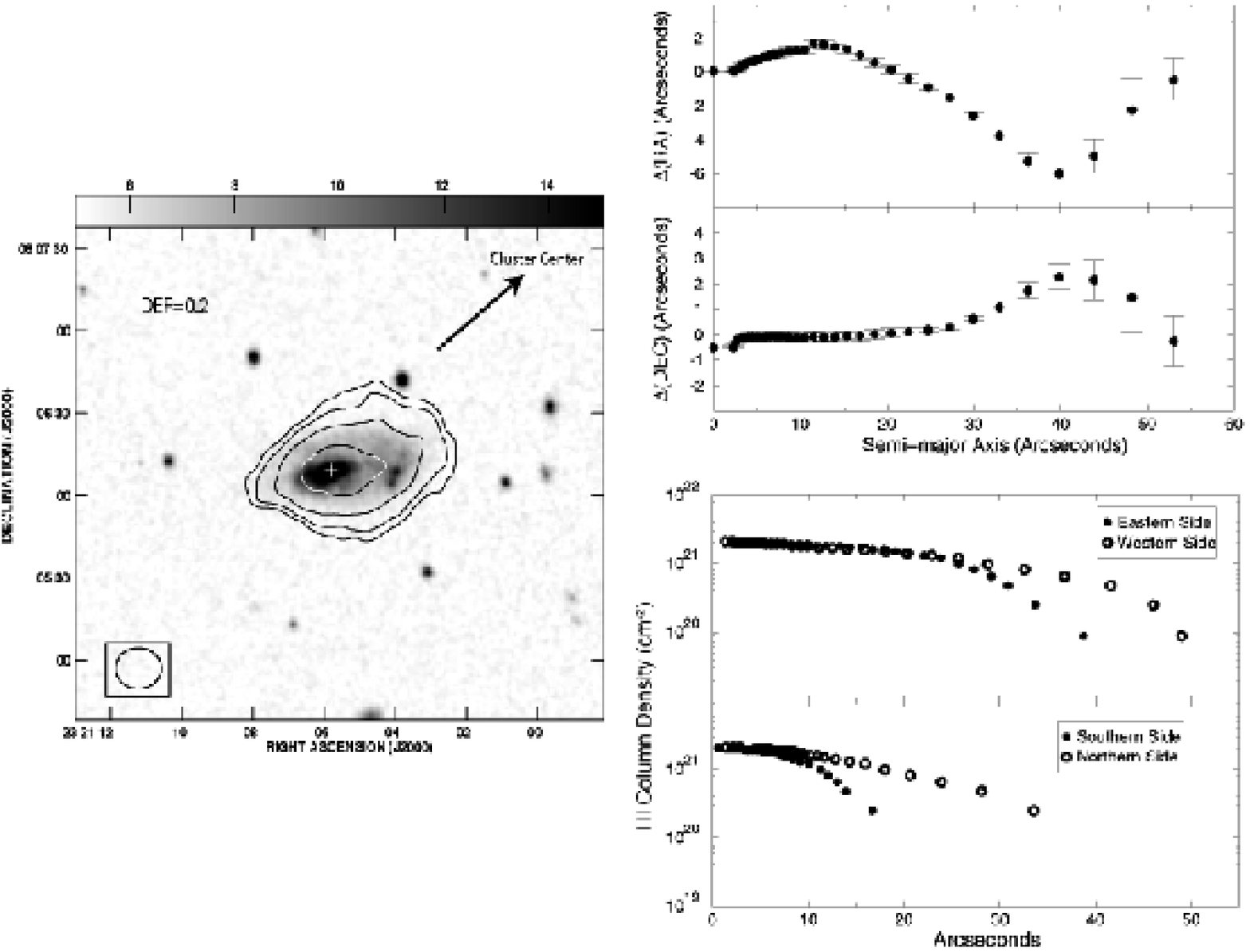}
\caption{\textbf{Left:} HI contours overlaid on DSS for KUG2318+078. There
  appears to be a slight asymmetry in the HI contours towards the
  NW. 
The lowest contour is 50
  mJy beam$^{-1}$ \kms which corresponds to 2.6x10$^{20}$
  cm$^{-2}$. The contour levels are at 2.6, 5.3, 11.1, and
  21.2x10$^{20}$ cm$^{-2}$.
\textbf{Top Right:} The 
  position center shifts of the ellipses fitted to the HI column
  density levels in KUG2318+078, measured in
  arcseconds, are plotted as a function of the semi-major axis. There
  is an HI 
  displacement towards the West and North. \textbf{Bottom Right:} The radial
  column density 
  distribution for KUG2318+078 along the major axis (top) and the minor
  axis (bottom). For the major axis, the filled circles represent the
  distribution from the galaxy center towards the East, and the
  empty circles represent the distribution from the galaxy center
  towards the West. For the minor axis, the filled circles
  represent the distribution towards the South, and the empty
  circles represent the distribution towards the
  North.\label{kug_comp}} 
\end{figure}

\begin{figure}
\plotone{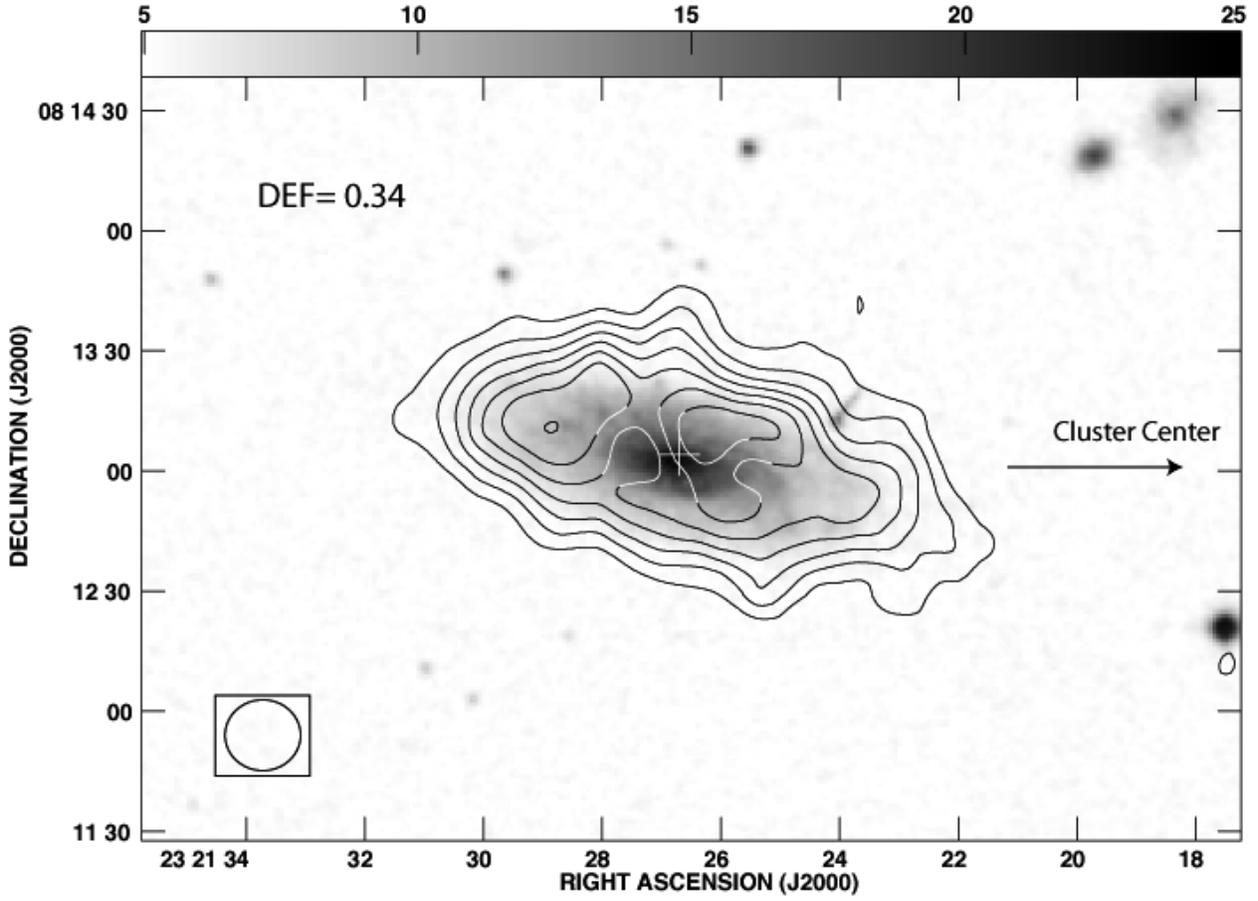}
\figcaption{HI contours overlaid on DSS for NGC7631.
The lowest contour is 50
  mJy beam$^{-1}$ \kms which corresponds to 2.2x10$^{20}$
  cm$^{-2}$. The contour levels are at 2.2, 4.3, 6.5, 8.7, 11.2, 13.1, and
  15.1x10$^{20}$ cm$^{-2}$.
\label{n7631_vla}}
\end{figure}

\begin{figure}[htbp]
\plotone{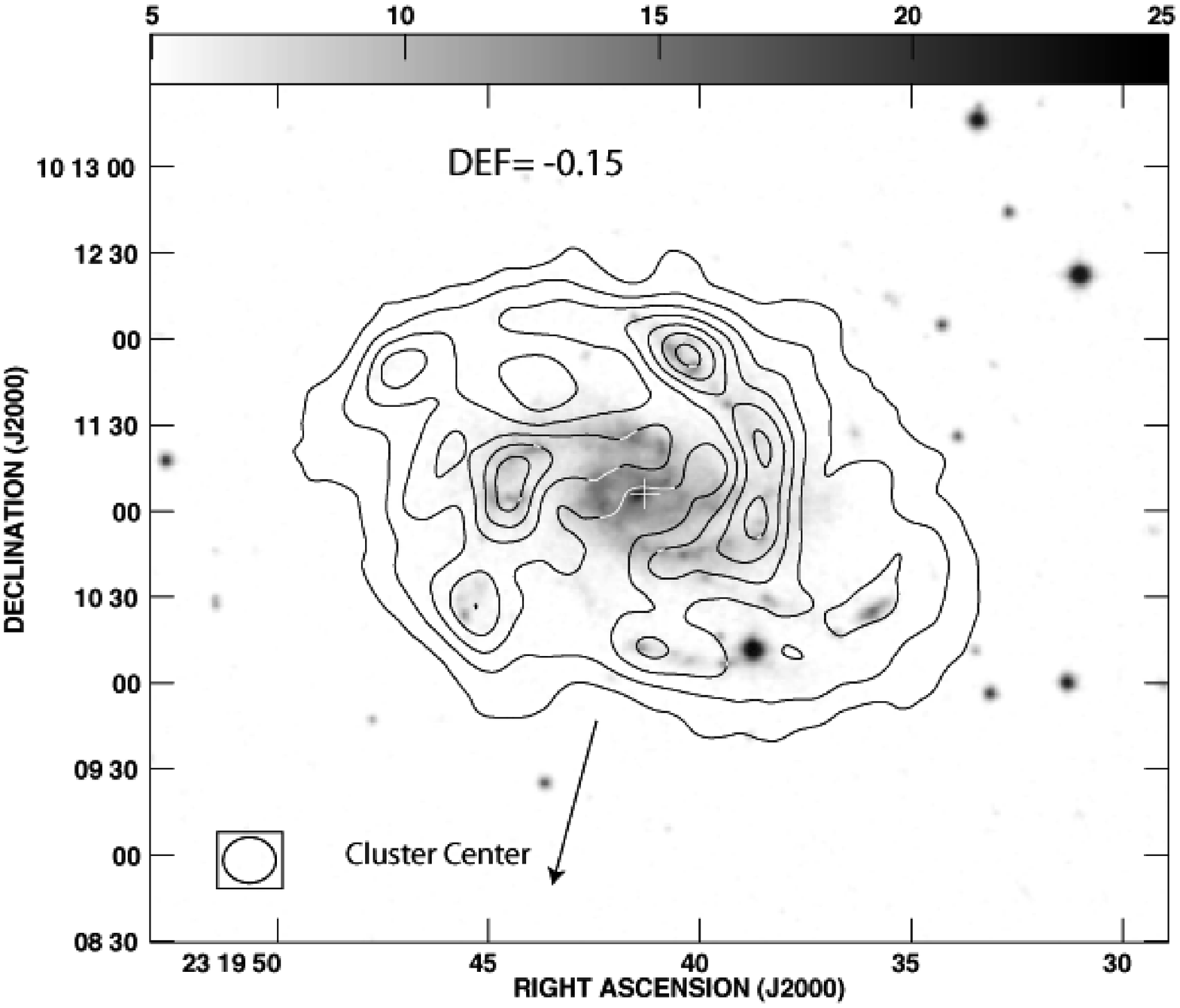}
\figcaption{HI contours overlaid on DSS for NGC7610.
The lowest contour is 100
  mJy beam$^{-1}$ \kms which corresponds to 4.5x10$^{20}$
  cm$^{-2}$. The contour levels are at 4.5, 9.0, 14.1, 18.0, 23.4, and
  27.2x10$^{20}$ cm$^{-2}$.
\label{n7610_vla}}
\end{figure}

\begin{figure}
\plotone{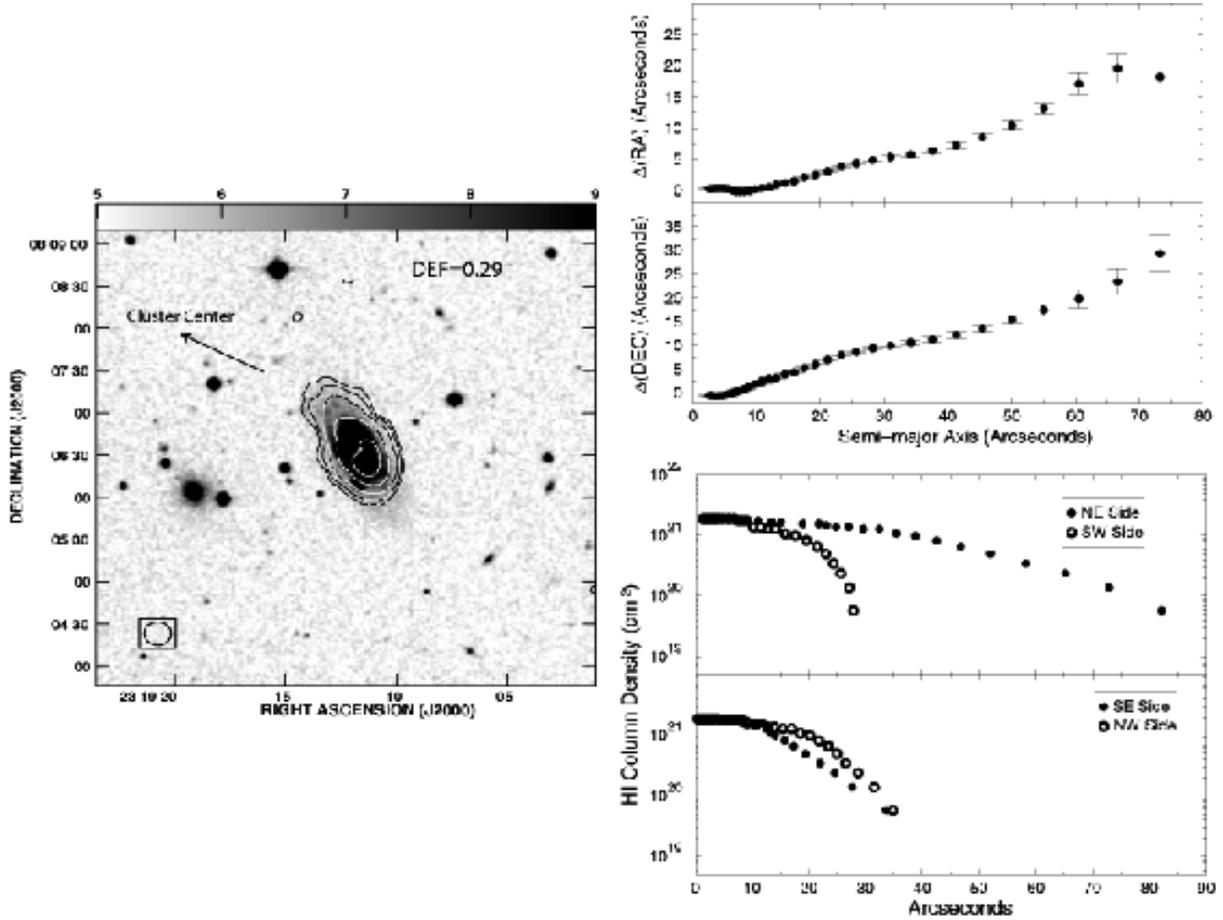}
\caption{\textbf{Left:} HI contours overlaid on DSS for IC5309. Note
  the HI disk is displaced from the optical counterpart and the
  asymmetry in the HI along the NE. 
The lowest contour is 30
  mJy beam$^{-1}$ \kms which corresponds to 1.4x10$^{20}$
  cm$^{-2}$. The contour levels are at 1.4, 2.8, 5.6, 11.3 and
  17.4x10$^{20}$ cm$^{-2}$.
\textbf{Top Right:} The 
  position center shifts of the ellipses fitted to the HI column
  density levels in IC5309, measured in
  arcseconds, are plotted as a function of semi-major axis. There is a
  large shift in the ellipse 
  position centers corresponding to an HI
  displacement towards the NE. \textbf{Bottom Right:} The radial
  column density 
  distribution for IC5309 along the major axis (top) and the minor
  axis (bottom). For the major axis, the filled circles represent the
  distribution from the galaxy center towards the North-East, and the 
  empty circles represent the distribution from the galaxy center
  towards the South-West. For the minor axis, the filled circles
  represent the distribution towards the South-East, and the empty
  circles represent the distribution towards the
  North-West. Note the extended HI gas along the NE
  side. \label{ic5309_comp}}  
\end{figure}

While the overall HI deficiencies observed with the Arecibo radio
telescope indicate that the ISM is being depleted in many spirals in
the Pegasus I 
cluster, a more direct indication as to the effects of ram pressure
stripping can be obtained from spatially resolved observations of the ISM.
Specifically, highly truncated HI disks have been found in both Virgo
 \citep{w88,ca90,ca94} and Coma \citep{ba01} cluster spirals, which
more directly indicates that a sweeping mechanism has depleted the outer
parts of the HI disks.  Perhaps even more telling are cases in which the
HI disk is significantly offset and/or distorted in a way that indicates
that much of the gas is now extraplanar \citep{k04,c05}.  This
decoupling between gas and stars almost surely implies a stripping event
caught in the act. As well, cluster spirals are found with extraplanar
radio continuum emission \citep{g95}.  To evaluate whether such
activity can be seen in the lower-richness environment of Pegasus I,
along with the Arecibo HI profiles and the optical imaging, we have
 obtained VLA HI images for 10 of the spiral
galaxies. 
Each of the disk galaxies for which we have HI imaging 
are
shown in Figures~\ref{n7608_comp} through~\ref{ic5309_comp}. 
For each of the HI contour images, the 
direction to the cluster center, defined by the central ellipticals,
is marked with an arrow. The optical center of the galaxy, as defined
by NED, is indicated with a cross. 
The position-velocity plots and the channel
maps for all of these galaxies can be viewed at
http://www.physics.unc.edu/$\backsim$llevy/pegasus. The total HI maps were made
by 
generating images of the total emission using the AIPS MOMNT task,
being careful to include only channels with line emission. MOMNT
smoothes and averages data in the three coordinate and velocity
dimensions. It is important to note that the HI fluxes obtained using
the VLA data and the fluxes obtained using the Arecibo data agree to
within 10\%.


The total HI map for NGC7608, is shown in
Figure~\ref{n7608_comp}. NGC7608 is an HI deficient galaxy with a
deficiency factor of 0.48 (Column 9 of Table~\ref{two}).
The most striking feature of this image is that the
HI disk is less extended than the optical disk, an unusual phenomenon
found only in cluster galaxies, and also attributed to ram pressure
stripping of the HI disk. Also note 
how the HI disk is asymmetric and displaced (in the SW direction)
with respect to its optical counterpart.

To quantify the asymmetry of the HI disk, elliptical contours were fit
to the HI surface density (specifically, the column density in
atoms/cm$^{2}$) using the IRAF ELLIPSE routine. The ELLIPSE routine
gave us the position centers of the fitted ellipses, in addition to
the HI column density, as a function of the radial distance. 
The top right of Figure~\ref{n7608_comp} shows the shift in the RA and
DEC centers (in arcseconds) of the fitted ellipse as a function of
distance along the semi-major axis. The ellipse center is seen to
shift towards the West and South at larger semi-major axis, in
accordance with the visual impression obtained from the HI contours
overlaid on the DSS, that the HI is offset from the
optical light in the South and 
West directions. From the fitted ellipses we then reconstruct the
radial HI column density profiles along both sides of the major and
minor axes. 
The bottom right of Figure~\ref{n7608_comp}
shows the radial profile of 
the HI distribution along the major and minor axis, which are both
clearly asymmetric. Specifically, the radial
profile along the major axis has a
steeper gradient on the NE side, while along the minor axis the
gradient is steeper on the SE side. To obtain a quantitative value
for the observed asymmetry, we calculated the skewness of the profiles on
both major and minor, with the result that the skewness is 1.1 and 0.6
along the major and minor axes respectively. In short, the systematic
shift in the HI position center with column density
level, and the associated skewness in the HI, provide a quantitative 
verification that the HI disk is displaced from the center of the optical
disk, and the displacement increases at fainter HI contour levels.
It is important to note that displaced HI disks
and asymmetric HI distributions are common in ``field'' galaxies (as
in the case 
of M101 \citep{a73}), and are thus not necessarily indicative of a ram
pressure 
event. However, as is more fully discussed in \S7, such displacements
between HI and optical, when coupled with other evidence such as the
high DEF values, and the 
presence of truncated HI disks, are further suggestive of an ISM-ICM
interaction.

\placefigure{n7608_comp_new.eps}

NGC7604 is one of the morphologically peculiar galaxies discussed in
Section~\ref{pecs}. Defining NGC7604 to be an Sa galaxy, we find it
has a DEF of 0.43. The total HI map, overlaid on the DSS optical
image, is displayed in
Figure~\ref{n7604_vla}, while Figure~\ref{n7604ha_vla} shows the HI
contours overlaid on the H$\alpha$ image obtained at the MDM telescope
(see \S{\ref{pecs}}). 
It is important to note that the higher velocity gas that
extends out past 3800 \kms in the Arecibo profile (Fig.~\ref{n7604})
is not seen in the 
VLA data. Evidently, that gas is too low in HI column density to be
detected by the VLA. 
The fact that the HI is concentrated to the SW side
of the galaxy is suggestive of an ISM-ICM interaction.  We discuss the
case of NGC7604 further 
in \S6, in the context of broadband and H$\alpha$ images.

\placefigure{n7604_vla_new.eps}
\placefigure{n7604ha_vla_new.eps}

Figure~\ref{u12480_vla} shows the total HI map of UGC12480, a low surface
brightness galaxy that has a normal HI
abundance, and the HI contours show a well behaved HI gas. UGC12480
was detected in the pointing of NGC7604, 
resulting in a higher rms noise after correction for the primary
beam. The small velocity gradient seen along the major axis of
the position-velocity plot (go to
http://www.physics.unc.edu/$\backsim$llevy/pegasus) implies that this galaxy is
seen close to face-on.  

\placefigure{u12480_vla_new.eps}

The HI map for Z406-042 is shown in Figure~\ref{z406_vla}. Z406-042 is
a deficient galaxy, with a deficiency factor of 0.41 (Column 9
of Table~\ref{two}).
The HI contours (Figure~\ref{z406_vla}) show a well behaved gas.

\placefigure{z406_vla_new.eps}

NGC7615 is a highly HI deficient galaxy, with a deficiency factor of
0.85. The HI contours for
NGC7615 (Figure~\ref{n7615_comp}) appear to be slightly
displaced from the  
optical disk with more gas appearing on the SE and SW sides. The measure of
skewness was calculated for NGC7615, using the same procedure used for
NGC7608, resulting in a skewness along the major axis of 0.07, and a
skewness along the minor axis of 0.37. This indicates there is little, if any,
skewness of the HI disk along the major axis. There is a small
asymmetry present along the minor axis, the HI gas is more extended
towards the SW. The top right panel of Figure~\ref{n7615_comp} shows
the 
ellipse position center shifts as a function of distance from the
galaxy center. The ellipse centers change very slightly, indicating a
small displacement of the HI gas towards the West and the South.
The bottom right panel of Figure~\ref{n7615_comp} shows the radial
profiles along the major and 
minor axis. Here we see the HI gas is mostly symmetric along the major
axis, with only a
slightly larger extent towards the SE edge. There is an asymmetry of
the HI gas along the minor axis, with more HI gas appearing towards
the SW edge. 

Though the asymmetry and displacement of the HI gas is small,
when coupled with a high deficiency factor and a truncated HI
disk, is indicative of a ram pressure event.

\placefigure{n7615_comp_new.eps}

Figure~\ref{u12535_vla} shows the HI contours of UGC12535, an HI
normal galaxy with a deficiency factor of 0.22. UGC12535 was
detected in the pointing of the central Pegasus cluster, resulting in
a higher value for rms noise after correction for primary
beam. Unfortunately, the 
VLA data, which had its velocity range centered for the central
Pegasus cluster, does
not cover the high  velocity range for UGC12535, though the low
velocity range is completely covered.
The channel map (http://www.physics.unc.edu/$\backsim$llevy/pegasus)
shows that 
the approaching side is in the 
NW. Figure~\ref{u12535_vla} shows that the HI disk is sharply cut off
in the NW 
to within the optical disk and it is asymmetrically placed with
respect to the disk. Since the HI velocity range is completely covered
in the NW, this is another example of a galaxy possibly affected by
the ICM ram pressure. Note that the cutoff in the SE is due to our
incomplete velocity coverage. 

\placefigure{u12535_vla_new.eps}

The HI contours for the HI normal galaxy KUG2318+078 are shown in 
Figure~\ref{kug_comp}. KUG2318+078 is also detected in the pointing of
the central Pegasus cluster. From the HI contours, there appears
to be more HI gas on the NW side.
The top right panel of Figure~\ref{kug_comp}
shows
ellipse position center shifts as a function of distance from the
galaxy center. The shift in ellipse centers indicate a displacement of
the HI gas towards the West and North.
The bottom right panel of Figure~\ref{kug_comp} also shows the radial
profiles along the major and 
minor axis. Here we see an asymmetry of the HI gas towards the West
along the major axis, with a skewness measure of 0.5, and an
asymmetry towards the North, along the minor axis, with a skewness
measure of 0.7. This asymmetry towards the NW is moderate, i.e., less
pronounced than in the case of NGC7608, but also indicative of
an asymmetric HI disk.

\placefigure{kug_comp_new.eps}

The total HI map for the slightly deficient galaxy NGC7631 is shown in
Figure~\ref{n7631_vla}. The HI gas is distributed fairly
symmetrically, with a slight amount of more extended gas on the
western edge. 

\placefigure{n7631_vla_new.eps}

NGC7610 is an HI normal (DEF=-0.15) disk galaxy. The HI contours are
shown in Figure~\ref{n7610_vla}, and we see symmetrically
distributed HI gas. 

\placefigure{n7610_vla_new.eps}

IC5309 is a moderately HI deficient galaxy, with a deficiency factor of
0.29. Figure~\ref{ic5309_comp} shows the HI contours overlaid on the
DSS image. The HI contours appear to be displaced from the optical
counterpart towards the NE, and similarly there appears to be an
asymmetry of the 
HI gas also favoring the NE side. The top right panel of
Figure~\ref{ic5309_comp} shows 
the ellipse position center shifts as a function of distance
from the galaxy center. The ellipse centers are shifting greatly,
indicating a displacement of 
the HI towards the East and North. The radial profile of the HI
distribution along 
the major and minor axis are also shown in Figure~\ref{ic5309_comp}. 
The radial profile along the South-West side has a much steeper gradient than
that for the North-East. The skewness was calculated along the major
and minor axis and found to be 1.1 and 0.1 respectively. There is a
large asymmetry of the HI gas along the major axis, with an extended
HI disk towards the North-East. 

\placefigure{ic5309_comp_new.eps}

\subsection{Evidence for Truncated HI Disks}
\begin{figure}[htb]
\plotone{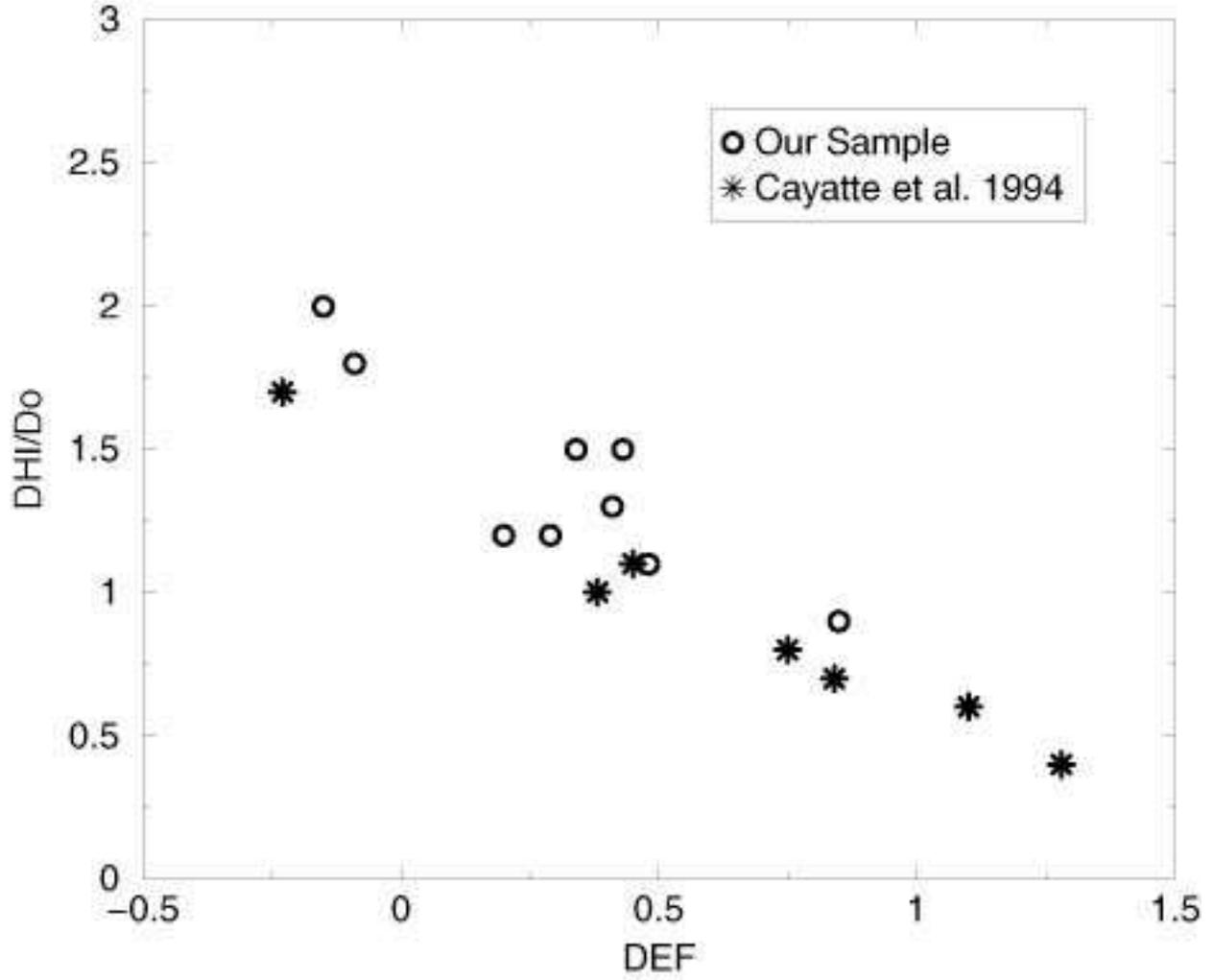}
\figcaption{HI to optical diameter ratios plotted versus HI
  deficiency. The circles represent our Pegasus cluster data
  and the asterisks represent data from the Virgo cluster taken from
  \citet{ca94}.\label{ratio}} 
\end{figure}

As mentioned earlier, truncated HI disks have been found in HI maps of
Virgo and Coma cluster spirals which also have large HI
deficiencies. \citet{ca94} carried out an extensive VLA survey of
spatially resolved HI for disk
galaxies in Virgo, as well as for a sample of field galaxies, to measure
the level of HI disk truncation in the 
Virgo cluster. Their results show that HI deficient galaxies have
truncated HI disks.
Having seen above some evidence for displaced (hence perhaps
extraplanar), and asymmetric
HI disks in a few HI deficient Pegasus I spirals (NGC7608, NGC7615,
KUG2318+078, and IC5309) we now turn our
attention to whether the disks are also truncated. 
Measurements of the diameter of the HI disks, D$_{HI}$, were made for
the 9 galaxies for which we have VLA data (assuming NGC7604 is a type
Sa galaxy), following the prescription of
\citet{ca94}, where the HI diameter is determined to be where the HI
column density equals 10$^{20}$ cm$^{-2}$. We also obtained optical
face-on diameters, D$_{o}$, corrected for extinction and inclination,
from \citet{b96}. In Figure~\ref{ratio} we have plotted the ratio
D$_{HI}$/D$_{o}$, between the HI diameter 
and the optical face-on diameter, against HI deficiency. The circles
represent our Pegasus cluster data, and the asterisks 
are galaxy data 
taken from \citet{ca94}. In agreement with the results of \citet{ca94},
we find that HI deficient galaxies tend to have smaller
D$_{HI}$/D$_{o}$ ratios. In addition, not only are the HI deficiencies in
Pegasus less severe than in Virgo, but the HI disk truncation is less
pronounced as well. Of the 9 Pegasus galaxies, 7 of these are Group I
galaxies (as defined by \citet{ca94}), and two (NGC7615 and Z406-042)
are Group II galaxies. In Group I galaxies only the outer parts of
the galaxies are depleted, and in Group II galaxies, there is more
depletion and a lower central column density. In comparison, Virgo
contains many Group III galaxies, which are thought to be undergoing
violent stripping and disk truncation.
To summarize at this point, not only are HI
deficient spiral galaxies  
found in the core of the Pegasus I cluster, and less frequently in the
foreground and background groups, but evidence is found as well for truncated
HI disks, and for HI disks that are offset from the optical (stellar) disks.
All of these effects have been observed at a higher level in Virgo
cluster spirals, and the effects are attributed there to ram pressure
stripping of the galaxies' ISM.

\placefigure{ratio_new.eps}

Table~\ref{vla_offsets} summarizes the results for the 9 galaxies
discussed in this section (UGC12535 is not included because the high
velocity range is completely missing from our data). The galaxy name
is given in the first column, followed by the HI deficiency in the
second column. Column 3 gives a measure of the displacement between
the HI and optical centers, normalized by the optical
diameter D$_{o}$. The displacement is
calculated at the point where the HI column density equals
5x10$^{20}$ cm$^{-2}$ 
and is simply the difference between the optical center, as
defined by NED, and the center of the fitted HI ellipse at that column
density. It is important to note that in the case of IC5309, there
is a large displacement between the HI and optical centers at the
highest HI column density, but at the level of 5x10$^{20}$ cm$^{-2}$,
there is only a small offset between the optical value and the HI
fitted ellipse 
center. For the rest of the galaxies, the displacement stays constant
between the highest HI column density and the column density at
5x10$^{20}$ cm$^{-2}$. Column 4 indicates whether or not there is a
shift in the HI center as a function of increasing semi-major
axis. NGC7615 has a question mark next to the yes because the shift is
only 2 arcsec, and the error is 1 arcsec. The other galaxies that show 
shifts show significant shifts in their HI centers, much larger than
the errors in the measurement. Column 5 indicates whether or not there
is an asymmetry in the HI contours. The four galaxies which show an
asymmetry are the ones for which the position center
shifts of the fitted ellipses, and HI profiles are shown in the
previous section. Column 6 gives a measure of HI disk truncation. In
order to obtain this value, a least squares fit was made to
Figure~\ref{ratio}. From the fit, the value of D$_{HI}/D_{o}$ at
DEF=0.0 was found to be 1.7. The HI disk truncation is the difference
between the measured value of D$_{HI}/D_{o}$ and the fiducial value of
1.7. The highly deficient galaxies show a more negative value for this
measure of HI disk truncation.   
 
\placetable{vla_offsets}

\section{NGC7604 and NGC7648\label{pecs}}
\begin{figure}[htb]
\plotone{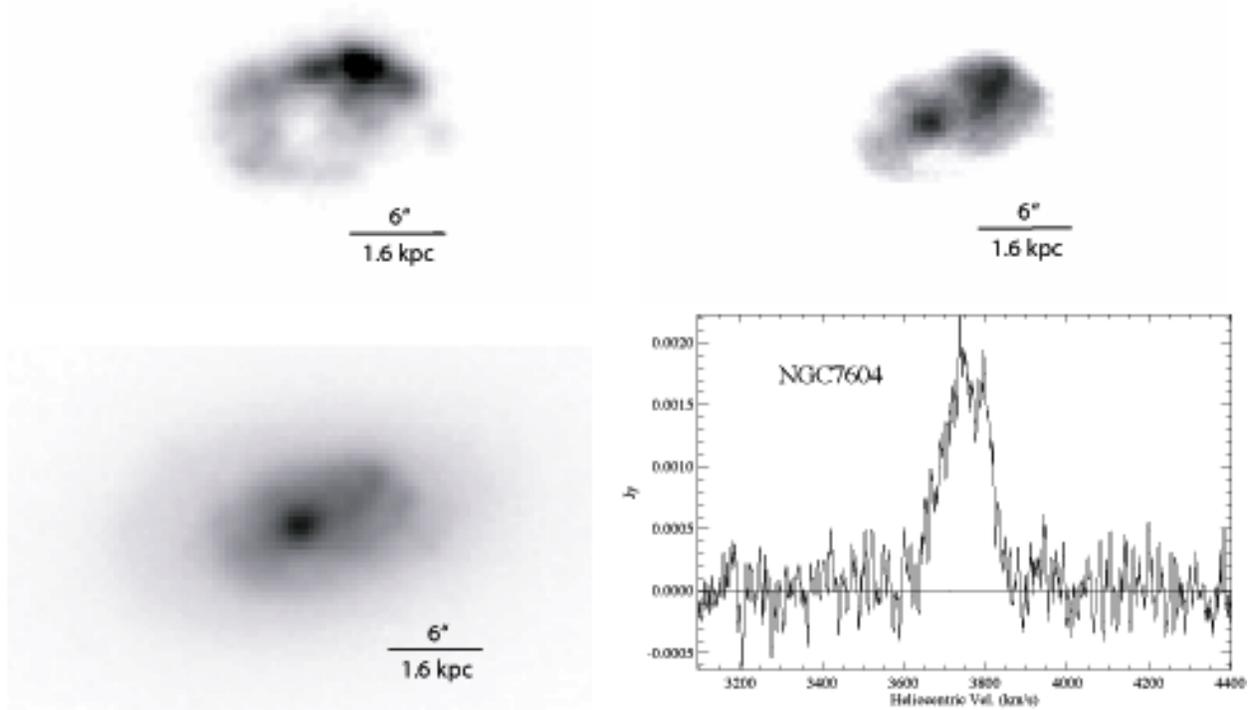}
\figcaption{\textbf{Top Left:} NGC7604 H$\alpha$ difference. The
      seeing is 1.2 arcseconds FWHM. The image has been oriented such
      that North is up and East is to 
      the left. Note the strong arc of star formation in the
      NW. \textbf{Top Right:} NGC7604 B-band. The seeing is 1.6
      arcseconds. Note the enhanced B-band 
      emission along the NW, coinciding with the H$\alpha$
      emission. \textbf{Bottom Left:} NGC7604 I-band. The seeing
      is 1.4 arcseconds. There is a well
      defined bulge and disk structure for the older stars. 
      \textbf{Bottom Right:} NGC7604 HI profile. \label{n7604}}
\end{figure}

\begin{figure}[htb]
\plotone{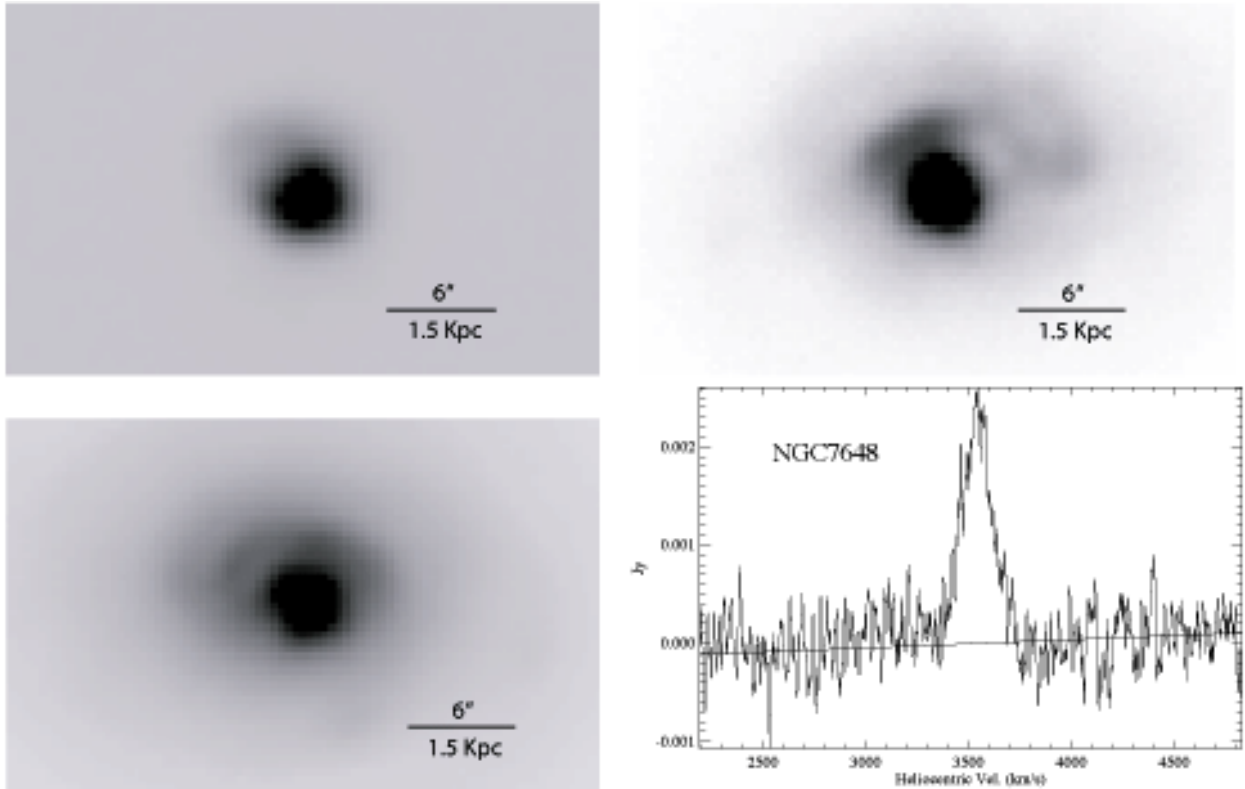}
\figcaption{\textbf{Top Left:} NGC7648 H$\alpha$ difference. 
      The seeing is 1.2 arcseconds FWHM.
      The image has been oriented such that North is up and East is to
      the left. There is centrally concentrated ongoing star formation.
      \textbf{Top Right:} NGC7648 B-band. The seeing is 2.1
      arcseconds. Note the asymmetric stellar ripples
      visible in the NE edge.
      \textbf{Bottom Left:} NGC7648 I-band. The seeing is 1.6
      arcseconds. The stellar ripples in the NE
      seen in the B-band are also visible in the I-band. 
      \textbf{Bottom Right:} NGC7648 HI profile.\label{n7648}}
\end{figure}

NGC7604 and NGC7648 are two morphologically peculiar galaxies whose
characteristics indicate different scenarios for their unusual
star formation (see Figures~\ref{n7604} and~\ref{n7648}).
The seeing in these images is
1-2 arcseconds FWHM. The H$\alpha$ image of NGC7604
(top left panel of Figure~\ref{n7604}) exhibits an asymmetric arc of
star formation 
concentrated along the NW edge, which coincides with the bright
off-nuclear emission region visible in the B-band image displayed in
top right panel of Figure~\ref{n7604}. In contrast, the I-band image, in
the bottom left panel of Figure~\ref{n7604} shows well defined bulge
and disk structures for 
the older stars. The arc of young star formation, concentrated on one
edge of the disk, is usually regarded as the signature of a ram
pressure event. 

Figure~\ref{n7604ha_vla} shows the HI contours overlaid on the
H$\alpha$ image. The bright arc of young
star formation in the NW corner, coincides with a region of HI
emission. This is surprising, since in a ram pressure stripping
scenario, one would expect the leading edge of the galaxy to have star
formation due to the ram pressure, but little, or no, HI gas in the
same area. In other words, the neutral gas gets pushed out by the ram
pressure, and we expect the HI gas to lie on the trailing edge,
opposed to the H$\alpha$. Figure~\ref{n7604ha_vla} also shows a
truncated H$\alpha$ disk, with respect to the HI disk. The HI disk
extends to a diameter of approximately 30 arcseconds (7 kpc), while
the H$\alpha$ diameter is only about 15 arcseconds (3.5
kpc). Truncated H$\alpha$ disks are commonly seen in galaxies
undergoing ram pressure stripping, as is the case of many Virgo
cluster disk galaxies \citep{kk04,c05,k04}.

On the other hand, NGC7648 shows centrally concentrated ongoing star
formation (top left panel of Figure~\ref{n7648}), and asymmetric
stellar
ripples visible in 
both the B-band (top right panel of Figure~\ref{n7648}) and I-band
(bottom left panel of Figure~\ref{n7648}) 
images. Further views of the ripple structures in NGC7648 can be seen in
\citet{r01}. The Arecibo HI profile shows the single peak structure that
indicates centrally concentrated gas.  The centrally concentrated HI gas and
star formation, in combination with the outer ripple structure
indicate a recent tidal interaction. Thus, 
NGC7604 and NGC7648 indicate that both
scenarios of galaxy-galaxy interactions and galaxy-ICM interactions are
likely to be operating in the Pegasus I cluster.

\placefigure{n7604_mdm_new.eps}
\placefigure{n7648_mdm_new.eps}

\section{Discussion}
Our principal observational conclusions at this point can be summarized as
follows:\\
\noindent 1) Approximately 40\% of the spiral galaxies in the core of
the Pegasus I 
cluster show mild HI deficiencies of about a factor of 2.  In contrast, in the
denser environment of the Virgo cluster, HI deficiencies are typically
a factor of 4, or greater.

\noindent 2) Outside of the core of Pegasus I, i.e., in the foreground and
background groups, little evidence is seen for HI depletion in the spirals.

\noindent 3) There is evidence as well that the HI deficient galaxies in
Pegasus I also have mildly truncated gas disks.

\noindent 4) Some evidence has been found for gas removal in progress in
specific galaxies, in the form of HI disks that are offset from the
optical disks, indicating that the gas is being displaced from the stars.

\noindent 5) In all respects the gas depletion effects are substantially
smaller than that seen in Virgo cluster spirals.

The results of our analysis, then, appear to point towards similar gas
depletion effects occurring in Pegasus I as are found in Virgo, just smaller
in magnitude.  However, as was mentioned at the start of \S4, simple ram
pressure arguments indicate that while ram pressure stripping should be
effective in the centers of major clusters such as Virgo and Coma, where
the ICM density is high and the galaxy velocity dispersion is also high, in
the case of Pegasus I, the basic ram pressure argument fails by nearly 2
orders of magnitude, even in the cluster center.  Thus while observationally
there appears to be only a gradual decline in the effectiveness of ram
pressure stripping with decreasing environmental density, this result is
at odds with the \citet{gg72} ram pressure argument, which predicts
complete stripping at any galactic radius for which the ram pressure
on the HI exceeds 
the gravitational restoring force, or none if this condition is not met.
Consequently, one must seriously reconsider whether the ram pressure
mechanism is really at work in Pegasus I, or whether 
the \citet{gg72} stripping criterion is overly simplistic, and we are
witnessing a partial stripping of the (multi-phase) HI gas.  Before
considering 
this further, we first summarize other observations which indicate stripping
effects in surprisingly low density environments.  

\citet{d97} obtained optical R band and H$\alpha$ images, as well as
X-ray data of NGC2276 in the NGC2300 group. Although the
NGC2300 group has
a low-density ICM, much like Pegasus, and is a low velocity dispersion
group environment, NGC2276 appears
asymmetric
in the H$\alpha$ image, in a manner that is reminiscent
of a bow shock produced in a ram pressure event. 
However, the R band
image, which tracks the older stellar population, is asymmetric as well,
which is not expected in a ram pressure event affecting only the gas.
Thus, given as well the fact that the calculated ram pressure is low, 
\citet{d97} conclude that a tidal disturbance has taken place. A similar 
conclusion is reached in the case of NGC4273 in the NGC4261 group, i.e., an
asymmetric spiral galaxy in a rather low-density environment.
Finally, NGC4522 in 
the Virgo cluster presents another special case \citep{k04,v00,kk99}.
  While the ram pressure in the center
of the Virgo cluster should be high enough to strip the ISM from disk
galaxies, NGC4522 is located beyond the radius of strong X-ray emission, and 
the ram pressure force at this radius is 10 times smaller than the force
needed to strip the gas.  Nevertheless, optical broadband and H$\alpha$
imaging, as well as resolved 21 cm and radio continuum data, all point towards
the conclusion that NGC4522 is {\it currently} in the act of being
stripped of  
its ISM, given the level of extraplanar gas and radio continuum emission 
\citep{k04}.

In short, other studies indicate examples, in addition to the Pegasus I
cluster, in which spiral galaxies are being divested of their ISM under
circumstances in which the ram pressure fails by an order of magnitude, or
more, to provide sufficient pressure to strip the ISM.  It
is therefore necessary to explain these effects either by considering
other mechanisms, such as tidal interactions, or by
reformulating our idea of ram pressure stripping. While tidal
interactions and/or preprocessing may explain some of the observed
phenomena in many of these clusters \citep{v04,vg03,d97}, it fails to
explain observed characteristics in other clusters. In low density
environments, such as Pegasus, where the observed HI deficiencies are
moderate at around factors of 2, (as opposed to Virgo and Coma, where
HI deficiencies can reach factors of 8 and more) it may be necessary
to reconsider ram pressure stripping. In dense environments, ram
pressure is capable of affecting the ISM as a whole, stripping away
the HI gas and leaving the disk galaxies with severely truncated HI
disks. It may be that in low density environments, ram pressure can
operate on the lower density component of the galaxy's ISM, without
being able to completely strip and 
disrupt the entire HI disk.

This work was partially supported by NSF grant AST-0406443 to the
University of North Carolina and by an NSF grant AST-0607643 to
Columbia University. We would like to thank the referee for the
valuable feedback and insight provided.

\clearpage

\clearpage
\begin{landscape}
\begin{deluxetable}{lcccccccc}
\tabletypesize{\scriptsize}
\tablewidth{0pc}
\tablecaption{VLA Instrumental Parameters\label{vla_obs}}
\tablehead{
  \colhead{Parameter} &\colhead{NGC7604} &\colhead{NGC7615}
  &\colhead{Z406-042} &\colhead{NGC7608}
  &\colhead{NGC7631} &\colhead{NGC7610} &\colhead{Pegasus
    Center}&\colhead{IC5309}\\ 
}
\startdata
\multicolumn{9}{l}{Phase Center}\\
RA (2000): (h m s)& 23 17 51.9& 23 19 54.4& 23 17 05.5& 23 19 15.3& 23
  21 26.7& 23 19 41.4& 23 20 32.1 & 23 19 11.6\\ 
DEC (2000): (d ' '')& 07 25 48.0& 08 23 58.0& 07 07 22.0& 08 21 01.0&
08 13 04.0& 10 11 06.0& 08 11 26.4 & 08 06 34.0\\
Velocity Center: (km/s)&3782&3071&3564&3508&3754&3354&3650&4198 \\
Velocity Range: (km/s)&600&600&600&600&600&600&1200&600\\
Time on Source: (hrs)&7&7&7&7&7&7&3&7\\
Bandwidth: (MHz)&3.125&3.125&3.125&3.125&3.125&3.125&6.25&3.125\\
Number of Channels&63&63&63&63&63&63&63&63\\
Velocity Resolution: (km/s)&10.6& 10.5& 10.5& 10.5& 10.6& 10.5&
21.1&10.6\\ 
\multicolumn{9}{l}{Synthesized Beam (FWHM):}\\
(arcsec) &17.3x15.8&17.6x16.1&17.6x16.1&17.3x16.2&18.4x16.6&
17.9x16.3&18.5x13.6 &17.3x16.5\\
(position angle d)&4&-5&9&17.6&23&8.6&40&14.3\\
rms noise: (mJy/beam)&0.33&0.33&0.35&0.31&0.37&0.35&0.5&0.35\\
rms noise:&1.4&1.3&1.4&1.2&1.4&1.4&4.5&1.4\\
\multicolumn{9}{l}{(10$^{19}$ cm$^{-2}$ per velocity channel)}\\
\enddata 
\end{deluxetable}
\end{landscape}

\clearpage
\begin{deluxetable}{cccccccccc}
\tablewidth{0pc}
\tablecaption{Galaxy Properties\label{one}}
\tablehead{
  \colhead{(1)}&\colhead{(2)}&\colhead{(3)}&\colhead{(4)}&\colhead{(5)}&
  \colhead{(6)}&
  \colhead{(7)}&\colhead{(8)}&\colhead{(9)}&\colhead{(10)} \\
  \colhead{Name}&\colhead{UGC Name}&\colhead{RA}&
  \colhead{DEC}& 
  \colhead{V}& \colhead{T\tablenotemark{1}}&
  \colhead{m$_{o}$\tablenotemark{2}}&
  \colhead{a\tablenotemark{3}}& \colhead{b/a}& \colhead{Delta V} \\
}
\startdata
\multicolumn{10}{c}{Foreground Cluster} \\
\\
\nodata &    UGC12361 &  23 06 22.4 &  11 17 08 &  2992 &  10$^{a}$
&15.60$^{a}$ & 1.0$^{a}$ & 0.400 & 184 \\  
NGC7537 &    UGC12442 &  23 14 34.5 &  04 29 55 &  2674 &   4$^{a}$
&12.72$^{a}$ & 2.1$^{a}$ & 0.238 & 368 \\  
NGC7541 &    UGC12447 &  23 14 43.9 &  04 32 04 &  2678 &  10$^{a}$
&11.57$^{a}$ & 3.4$^{a}$ & 0.324 & 487 \\  
\nodata &    UGC12522 &  23 20 16.6 &  08 00 20 &  2812 &   9$^{a}$
&15.19$^{a}$ & 1.7$^{a}$ & 0.941 & 126 \\  
\nodata &    UGC12544 &  23 21 45.1 &  09 04 40 &  2844 &  10$^{a}$
&14.56$^{a}$ & 1.2$^{a}$ & 0.917 &  84 \\  
\nodata &    UGC12580 &  23 24 33.8 &  08 36 58 &  3009 &   1$^{a}$
&16.34$^{b}$ & 1.3$^{a}$ & 0.231 & 220 \\  
NGC7615 &    \nodata  &  23 19 54.4 &  08 23 57 &  3071 &   3$^{a}$
&14.48$^{a}$ & 1.0$^{b}$ & 0.500 & 250 \\  
\\
\multicolumn{10}{c}{Central Cluster} \\
\\
\nodata &    UGC12304 &  23 01 08.3 &  05 39 16 &  3470 &   5$^{b}$
&13.49$^{a}$ & 1.4$^{a}$ & 0.143 & 322 \\  
\nodata &    UGC12382 &  23 07 55.2 &  05 09 40 &  3523 &   6$^{a}$
&14.74$^{a}$ & 1.2$^{a}$ & 0.083 & 288 \\  
 IC1474 &    UGC12417 &  23 12 51.2 &  05 48 23 &  3506 &   6$^{a}$
 &14.00$^{a}$ & 1.1$^{a}$ & 0.455 & 284 \\  
NGC7518 &    UGC12422 &  23 13 12.8 &  06 19 18 &  3536 &   1$^{a}$
&13.81$^{a}$ & 1.5$^{a}$ & 0.933 &  83 \\  
\nodata &    UGC12451 &  23 14 45.5 &  05 24 55 &  3645 &  10$^{a}$
&14.86$^{a}$ & 1.6$^{a}$ & 0.250 & 200 \\  
NGC7563 &    UGC12465 &  23 15 55.9 &  13 11 46 &  4174 &   1$^{a}$
&13.43$^{a}$ & 2.1$^{a}$ & 0.429 & 300 \\  
\nodata &    UGC12467 &  23 16 01.4 &  06 39 08 &  3507 &   8$^{a}$
&14.40$^{a}$ & 1.5$^{a}$ & 0.267 & 215 \\  
\nodata &    UGC12480 &  23 17 27.3 &  07 37 55 &  3872 &   9$^{a}$
&17.01$^{b}$ & 1.0$^{a}$ & 1.000 & 115 \\  
NGC7593 &    UGC12483 &  23 17 57.0 &  11 20 57 &  4108 &   3$^{b}$
&13.83$^{a}$ & 1.0$^{a}$ & 0.500 & 270 \\  
\nodata &    UGC12494 &  23 18 52.6 &  06 52 38 &  4196 &   7$^{a}$
&14.39$^{a}$ & 1.5$^{a}$ & 0.333 & 233 \\  
\nodata &    UGC12497 &  23 19 10.8 &  07 42 13 &  3761 &  10$^{a}$
&14.90$^{a}$ & 1.1$^{a}$ & 0.273 & 188 \\  
 IC5309 &    UGC12498 &  23 19 11.7 &  08 06 34 &  4198 &   3$^{a}$
 &13.87$^{a}$ & 1.5$^{a}$ & 0.400 & 300 \\  
NGC7608 &    UGC12500 &  23 19 15.3 &  08 21 01 &  3508 &   3$^{b}$
&13.92$^{a}$ & 1.5$^{a}$ & 0.267 & 310 \\  
NGC7610 &    UGC12511 &  23 19 41.3 &  10 11 06 &  3554 &   6$^{a}$
&13.25$^{a}$ & 2.7$^{a}$ & 0.815 & 286 \\  
\nodata &    UGC12535 &  23 21 01.6 &  08 10 46 &  4214 &   4$^{a}$
&16.61$^{b}$ & 1.1$^{a}$ & 0.182 & 215 \\  
NGC7631 &    UGC12539 &  23 21 26.7 &  08 13 03 &  3754 &   3$^{a}$
&13.12$^{a}$ & 1.8$^{a}$ & 0.444 & 385 \\  
\nodata &    UGC12553 &  23 22 13.7 &  09 23 03 &  3573 &   9$^{a}$
&17.00$^{a}$ & 1.4$^{a}$ & 0.786 & 102 \\  
\nodata &    UGC12561 &  23 22 58.5 &  08 59 37 &  3743 &   8$^{a}$
&14.93$^{a}$ & 1.7$^{a}$ & 0.235 & 217 \\  
\nodata &    UGC12562 &  23 22 47.3 &  11 46 22 &  3836 &   8$^{a}$
&16.61$^{b}$ & 1.3$^{a}$ & 0.231 & 181 \\  
NGC7643 &    UGC12563 &  23 22 50.4 &  11 59 20 &  3878 &   5$^{b}$
&13.61$^{a}$ & 1.4$^{a}$ & 0.571 & 349 \\  
\nodata &    UGC12571 &  23 23 22.5 &  13 19 09 &  3913 &   3$^{b}$
&14.32$^{a}$ & 2.0$^{a}$ & 0.550 & 303 \\  
\nodata &    UGC12585 &  23 24 39.6 &  08 25 32 &  3675 &   8$^{a}$
&14.39$^{a}$ & 1.6$^{a}$ & 0.938 & 115 \\  
Z406-042 &   \nodata  &  23 17 05.5 &  07 07 22 &  3564 &   5$^{b}$
&15.02$^{a}$ & 0.9$^{a}$ & 0.667 & 223 \\  
Z406-054 &   \nodata  &  23 18 16.2 &  06 49 32 &  3428 &   1$^{a}$
&15.80$^{c}$ & 1.0$^{b}$ & 0.200 & 206 \\  
KUG2318+078& \nodata  &  23 21 05.8 &  08 06 09 &  3886 &   4$^{b}$
&14.49$^{a}$ & 1.1$^{b}$ & 0.455 & 182 \\  
Z406-086 &   \nodata  &  23 21 40.9 &  08 59 24 &  3606 &  10$^{b}$
&14.41$^{a}$ & 1.3$^{b}$ & 0.462 & 208 \\  
OBC97p05-6 & \nodata  &  23 21 47.0 &  09 02 26 &  3667 &   4$^{a}$
&16.97$^{a}$ & 0.8$^{b}$ & 0.875 & 129 \\  
FGC284A &    \nodata  &  23 22 58.5 &  07 40 20 &  3471 &   5$^{b}$
&17.70$^{d}$ & 1.1$^{b}$ & 0.182 & 162 \\  
NGC7604 &    \nodata  &  23 17 51.8 &  07 25 49 &  3782 & pec$^{b}$
&15.27$^{a}$ & 0.3$^{b}$ & 0.667 & 175 \\  
NGC7648 &    UGC12575 &  23 23 54.1 &  09 40 04 &  3559 & pec$^{b}$
&13.42$^{a}$ & 1.6$^{a}$ & 0.625 & 277 \\  
\\
\multicolumn{10}{c}{Background Cluster} \\
\\
NGC7469 &    UGC12332 &  23 03 15.6 &  08 52 26 &  4892 &   1$^{a}$
&12.64$^{a}$ & 1.6$^{a}$ & 0.688 & 238 \\  
\nodata &    UGC12370 &  23 07 06.4 &  09 57 38 &  4892 &   6$^{a}$
&14.13$^{a}$ & 1.5$^{a}$ & 0.200 & 281 \\  
NGC7495 &    UGC12391 &  23 08 57.2 &  12 02 53 &  4887 &   5$^{a}$
&13.56$^{a}$ & 2.0$^{a}$ & 0.900 & 224 \\  
NGC7511 &    UGC12412 &  23 12 26.3 &  13 43 36 &  4928 &   2$^{b}$
&14.16$^{a}$ & 1.1$^{a}$ & 0.455 & 301 \\  
NGC7515 &    UGC12418 &  23 12 48.7 &  12 40 45 &  4475 &   5$^{b}$
&13.05$^{a}$ & 1.7$^{a}$ & 0.824 & 334 \\  
\nodata &    UGC12423 &  23 13 13.1 &  06 25 48 &  4839 &   5$^{a}$
&12.74$^{a}$ & 3.6$^{a}$ & 0.111 & 515 \\  
\nodata &    UGC12426 &  23 13 32.7 &  06 34 05 &  4720 &   6$^{a}$
&14.54$^{a}$ & 1.3$^{a}$ & 0.154 & 268 \\  
NGC7529 &    UGC12431 &  23 14 03.2 &  08 59 33 &  4538 &   7$^{b}$
&14.36$^{a}$ & 1.1$^{a}$ & 0.909 & 191 \\  
NGC7535 &    UGC12438 &  23 14 12.8 &  13 34 55 &  4604 &   7$^{a}$
&14.18$^{a}$ & 1.7$^{a}$ & 1.000 & 138 \\  
NGC7536 &    UGC12437 &  23 14 13.2 &  13 25 34 &  4697 &   4$^{a}$
&13.32$^{a}$ & 2.2$^{a}$ & 0.364 & 354 \\  
NGC7570 &    UGC12473 &  23 16 44.7 &  13 28 59 &  4698 &   1$^{a}$
&13.50$^{a}$ & 1.6$^{a}$ & 0.500 & 204 \\  
NGC7580 &    UGC12481 &  23 17 36.4 &  14 00 04 &  4432 &   4$^{b}$
&14.07$^{a}$ & 0.8$^{a}$ & 0.750 & 275 \\  
NGC7591 &    UGC12486 &  23 18 16.2 &  06 35 09 &  4956 &   4$^{a}$
&13.01$^{a}$ & 1.9$^{a}$ & 0.421 & 435 \\  
\nodata &    UGC12547 &  23 21 51.6 &  05 00 23 &  5113 &   5$^{b}$
&14.16$^{a}$ & 1.2$^{a}$ & 0.500 & 248 \\  
\nodata &    UGC12555 &  23 22 34.0 &  05 07 13 &  4915 &   6$^{a}$
&16.61$^{b}$ & 1.1$^{a}$ & 0.273 & 256 \\  
 IC5283 &    \nodata  &  23 03 18.0 &  08 53 37 &  4804 &  10$^{a}$
 &14.34$^{a}$ & 0.8$^{b}$ & 0.500 & 385 \\  
 IC5292 &    \nodata  &  23 13 47.2 &  13 41 14 &  4612 &   3$^{b}$
 &15.20$^{c}$ & 0.5$^{b}$ & 1.000 & 214 \\  
\\
\multicolumn{10}{c}{Non-Pegasus Spirals} \\
\\
KUG2358+128A&\nodata& 00 01 13.4&13 08 39&5461&5$^{b}$ 
&15.2$^{a}$&1.0$^{b}$&0.500& 400\\
\nodata   &UGC00011&00 03 21.5&22 06 11&4447&2$^{b}$
&15.0$^{a}$&1.1$^{a}$&0.727& 200\\
NGC7816   &UGC00016&00 03 48.8&07 28 43&5240&4$^{a}$
&13.3$^{a}$&2.0$^{a}$&1.000& 300\\
NGC7817   &UGC00019&00 03 58.9&20 45 08&2309&4$^{a}$
&11.6$^{a}$&4.0$^{a}$&0.275& 450\\
\nodata   &UGC00024&00 04 14.7&22 35 19&4442&6$^{a}$
&14.6$^{a}$&1.2$^{a}$&0.667& 200\\
\nodata   &UGC00076&00 08 49.2&24 32 25&4581&5$^{b}$
&15.2$^{b}$&1.1$^{a}$&0.273& 250\\
\nodata   &UGC00079&00 09 04.4&25 37 07&4345&6$^{b}$
&14.8$^{a}$&1.7$^{b}$&0.765& 230\\
NGC0041   &\nodata& 00 12 48.0&22 01 24&5949&3$^{a}$
&14.1$^{a}$&1.0$^{a}$&0.500& 300\\
NGC0052   &UGC00140&00 14 40.1&18 34 55&5392&3$^{b}$
&13.1$^{a}$&2.6$^{a}$&0.192& 550\\ 
\nodata   &UGC00144&00 15 26.8&16 14 07&5620&4$^{b}$
&14.3$^{b}$&1.0$^{a}$&0.300& 400\\
\nodata   &UGC00164&00 17 23.7&18 05 03&5443&4$^{a}$
&14.1$^{a}$&1.8$^{a}$&0.389& 300\\
\nodata   &UGC00168&00 18 10.6&18 17 32&5521&1$^{a}$
&15.3$^{b}$&1.2$^{a}$&0.250& 500\\
\nodata   &UGC00179&00 19 00.6&23 28 36&4485&6$^{a}$
&14.5$^{a}$&1.3$^{a}$&0.538& 300\\
IC1544    &UGC00204&00 21 17.5&23 05 27&5714&5$^{a}$
&14.1$^{a}$&1.4$^{a}$&0.643& 200\\
IC1546    &\nodata& 00 21 29.0&22 30 21&5820&5$^{b}$
&14.5$^{a}$&1.0$^{b}$&0.500& 230\\
\nodata   &UGC00228&00 23 56.7&24 18 20&5683&4$^{a}$
&14.3$^{a}$&1.3$^{a}$&0.692& 250\\
IC1552    &UGC00297&00 29 43.7&21 28 37&5600&5$^{b}$
&14.3$^{a}$&1.0$^{a}$&0.200& 350\\
\enddata
\tablenotetext{1}{(a) obtained using the RC3 catalogue, (b) measured
  by the authors using the Palomar Sky Survey prints.} 
\tablenotetext{2}{(a) corrected magnitude obtained
using the RC3 catalogue, (b) the uncorrected magnitude is obtained
using the UGC, and corrected  for  galactic
extinction, internal extinction, and redshift correction as prescribed
in \citet{b96}, (c)
uncorrected magnitude, obtained from the Zwicky catalog, and corrected
for galactic 
extinction, internal extinction, and redshift correction as prescribed
in \citet{b96}, (d) uncorrected magnitude, obtained from the Flat
Galaxy Catalogue, and corrected
for galactic 
extinction, internal extinction, and redshift correction as prescribed
in \citet{b96}. }
\tablenotetext{3}{(a) obtained using the UGC catalogue, (b) measured
  by the authors using the Palomar Sky Survey prints.}
\end{deluxetable}

\clearpage
\begin{deluxetable}{ccccccccc}
\tablewidth{0pc}
\tablecaption{Derived and Observed Quantities\label{two}}
\tablehead{
\colhead{(1)}&\colhead{(2)}&\colhead{(3)}&\colhead{(4)}&\colhead{(5)}&
 \colhead{(6)}&
 \colhead{(7)}&\colhead{(8)}&\colhead{(9)}\\  
\colhead{Name\tablenotemark{1}}&\colhead{UGC
 Name}&\colhead{T}&\colhead{t$_{exp}$}&\colhead{Flux}&
\colhead{log}&\colhead{log}&\colhead{log}&\colhead{DEF} \\

\colhead{}&\colhead{}&\colhead{}&\colhead{}&\colhead{(Jy km/s)}&
 \colhead{(M$_{HI}$)}&                                
 \colhead{(L$_{B}$)}&   
 \colhead{(D$_{o}^{2}$)}&
 \colhead{}
}       
\startdata
\multicolumn{9}{c}{Foreground Cluster} \\
\\
\nodata & UGC12361 & 10&10 &  2.9$\pm$0.1 &  9.05 &  9.11 &  2.13&  0.01 \\ 
NGC7537 & UGC12442 &  4& 5 & 18.3$\pm$0.2 &  9.84 &  9.67 &  2.78& -0.31 \\
NGC7541 & UGC12447 & 10& 5 & 29.1$\pm$0.5 & 10.00 & 10.33 &  3.19& -0.10 \\
\nodata & UGC12522 &  9& 5 &  4.7$\pm$0.1 &  9.25 &  9.28 &  2.59&  0.19 \\
\nodata & UGC12544 & 10& 5 &  4.7$\pm$0.2 &  9.25 &  9.34 &  2.29& -0.06 \\
\nodata$^{*}$&UGC12580&1&10&0.3$\pm$0.1 & 8.04 & 8.82 & 2.36&  1.11 \\
NGC7615 & \nodata  &  3& 5 &  0.5$\pm$0.1 &  8.31 &  9.28 &  2.13&  0.85 \\
\\
\multicolumn{9}{c}{Central Cluster} \\
\\
\nodata & UGC12304 &  5&10 &  2.1$\pm$0.1 &  8.89 &  9.10 &  2.42&  0.38 \\
\nodata & UGC12382 &  6& 5 &  4.9$\pm$0.2 &  9.27 &  9.01 &  2.29& -0.08 \\
 IC1474 & UGC12417 &  6& 5 &  4.0$\pm$0.1 &  9.18 &  9.37 &  2.21 & -0.05 \\
NGC7518 & UGC12422 &  1& 5 &  2.4$\pm$0.5 &  8.96 &  9.62 &  2.48&  0.27 \\ 
\nodata & UGC12451 & 10& 5 &  3.8$\pm$0.1 &  9.16 &  9.28 &  2.54&  0.24 \\
NGC7563&UGC12465&1&10&$\leqslant$0.1$\pm$0.1&7.58&9.58&2.78&$\geqslant$1.82\\ 
\nodata & UGC12467 &  8& 5 &  2.9$\pm$0.1 &  9.04 &  9.59 &  2.48&  0.32 \\
\nodata$^{*}$&UGC12480&9&5&  3.7$\pm$0.1 &  9.15 &  8.55 &  2.13& -0.09 \\
NGC7593 & UGC12483 &  3& 5 &  3.2$\pm$0.2 &  9.09 &  9.53 &  2.13&  0.07 \\
\nodata & UGC12494 &  7& 5 &  5.1$\pm$0.1 &  9.29 &  9.27 &  2.48&  0.06 \\
\nodata & UGC12497 & 10& 5 &  3.9$\pm$0.1 &  9.17 &  9.28 &  2.21& -0.04 \\
 IC5309 & UGC12498 &  3& 5 &  3.2$\pm$0.1 &  9.09 &  9.34 &  2.48 & 0.29 \\
NGC7608 & UGC12500 &  3& 5 &  2.1$\pm$0.1 &  8.89 &  9.19 &  2.48&  0.48 \\
NGC7610 & UGC12511 &  6& 5 & 22.1$\pm$0.1 &  9.92 &  9.60 &  2.99& -0.15 \\
\nodata & UGC12535 &  4&10 &  1.3$\pm$0.1 &  8.68 &  8.71 &  2.21&  0.51 \\
NGC7631 & UGC12539 &  3& 5 &  3.6$\pm$0.1 &  9.13 &  9.78 &  2.64&  0.34 \\
\nodata & UGC12553 &  9& 5 &  3.6$\pm$0.1 &  9.14 &  8.55 &  2.42&  0.16 \\
\nodata & UGC12561 &  8& 5 &  4.7$\pm$0.2 &  9.25 &  9.38 &  2.59&  0.19 \\
\nodata & UGC12562 &  8& 5 &  3.2$\pm$0.1 &  9.09 &  8.71 &  2.36&  0.16 \\
NGC7643 & UGC12563 &  5&10 &  1.1$\pm$0.2 &  8.61 &  9.51 &  2.42&  0.65 \\
\nodata & UGC12571 &  3& 5 &  7.9$\pm$0.1 &  9.48 &  9.31 &  2.73&  0.05 \\
\nodata & UGC12585 &  8& 5 &  7.6$\pm$0.2 &  9.46 &  9.33 &  2.54& -0.06 \\
Z406-042$^{*}$&\nodata& 5&5&0.9$\pm$0.1 &  8.52 &  9.17 &  2.04&  0.41 \\
Z406-054$^{*}$&\nodata & 1& 10 & 0.4$\pm$0.1 & 8.18 & 9.03 & 2.13&  0.84 \\ 
KUG2318+078$^{*}$&\nodata&4& 5 & 2.6$\pm$0.1 & 9.00 & 9.16 & 2.21& 0.20 \\
Z406-086$^{*}$ &\nodata  & 10& 5 & 2.0$\pm$0.1 & 8.88 & 9.34 & 2.36&0.37\\
OBC97p05-6$^{*}$ &\nodata & 4& 5 & 2.7$\pm$0.1 & 9.00 & 8.56 & 1.94&0.02\\ 
FGC284A$^{*}$ &  \nodata &  5& 5 & 1.3$\pm$0.1 & 8.70 & 8.27 & 2.21 &0.38\\  
NGC7604 & \nodata &pec&60& 0.2$\pm$0.1 & 7.97 &  9.36 &  1.09&\nodata\\  
NGC7648 &  UGC12575 & pec&40& 0.4$\pm$0.1 & 8.21 & 10.01 & 2.54&\nodata\\  
\\
\multicolumn{9}{c}{Background Cluster} \\
\\
NGC7469 & UGC12332 &  1& 5 &  1.2$\pm$0.4 &  8.65 & 10.12 &  2.54&  0.62  \\  
\nodata & UGC12370 &  6& 5 &  6.6$\pm$0.2 &  9.39 &  9.11 &  2.48& -0.04  \\ 
NGC7495 & UGC12391 &  5& 5 & 12.3$\pm$0.1 &  9.67 &  9.62 &  2.73& -0.13  \\
NGC7511 & UGC12412 &  2& 5 &  2.3$\pm$0.2 &  8.93 &  9.33 &  2.21&  0.14  \\
NGC7515 & UGC12418 &  5& 5 &  5.1$\pm$0.1 &  9.29 &  9.82 &  2.59&  0.13  \\
\nodata & UGC12423 &  5& 5 & 16.7$\pm$0.1 &  9.80 &  9.41 &  3.24&  0.18  \\
\nodata & UGC12426 &  6& 5 &  3.7$\pm$0.1 &  9.14 &  9.54 &  2.36&  0.11  \\
NGC7529 & UGC12431 &  7& 5 &  4.3$\pm$0.1 &  9.21 &  9.62 &  2.21& -0.08  \\
NGC7535 & UGC12438 &  7& 5 &  4.8$\pm$0.1 &  9.26 &  9.32 &  2.59&  0.18  \\
NGC7536 & UGC12437 &  4& 5 & 11.2$\pm$0.1 &  9.63 &  9.44 &  2.82& -0.07  \\
NGC7570 & UGC12473 &  1& 5 &  7.5$\pm$0.1 &  9.45 &  9.65 &  2.54& -0.19  \\
NGC7580 & UGC12481 &  4& 5 &  5.4$\pm$0.1 &  9.31 &  9.51 &  1.94& -0.29  \\
NGC7591 & UGC12486 &  4& 5 & 14.1$\pm$0.2 &  9.73 &  9.77 &  2.69& -0.25  \\
\nodata & UGC12547 &  5& 5 &  3.5$\pm$0.1 &  9.12 &  9.30 &  2.29&  0.03  \\
\nodata & UGC12555 &  6& 5 &  3.8$\pm$0.1 &  9.15 &  8.71 &  2.21& -0.03  \\ 
 IC5283 & \nodata  & 10& 5 &  1.9$\pm$0.6 &  8.85 &  9.40 &  1.94&  0.05  \\
 IC5292 & \nodata  &  3& 5 &  1.8$\pm$0.1 &  8.83 &  9.27 &  1.53& -0.05  \\ 
\\
\multicolumn{9}{c}{Non-Pegasus Spirals} \\
\\
KUG2358+128A&\nodata & 5& 5& 1.6$\pm$0.2& 8.78& 9.27& 2.39& 0.46 \\
\nodata&   UGC00011& 2& 5& 1.8$\pm$0.2& 8.83& 9.35& 2.29& 0.29 \\
NGC7816&   UGC00016& 4& 5& 8.6$\pm$0.1& 9.51&10.03& 2.96& 0.13 \\
NGC7817&   UGC00019& 4& 5&11.8$\pm$0.1& 9.65&10.71& 2.85&-0.07 \\
\nodata&   UGC00024& 6& 5& 4.3$\pm$0.2& 9.21& 9.51& 2.37& 0.05 \\
\nodata&   UGC00076& 5& 5& 3.3$\pm$0.2& 9.09& 9.27& 2.32& 0.09 \\
\nodata&   UGC00079& 6& 5& 5.1$\pm$0.1& 9.28& 9.43& 2.66& 0.21 \\
NGC0041&   \nodata & 3& 5& 1.7$\pm$0.2& 8.80& 9.71& 2.47& 0.56 \\
NGC0052&   UGC00140& 3& 5& 3.9$\pm$0.2& 9.16&10.11& 3.21& 0.66 \\
\nodata&   UGC00144& 4&15& 0.6$\pm$0.5& 8.36& 9.63& 2.42& 0.96 \\
\nodata&   UGC00164& 4& 5& 1.1$\pm$0.2& 8.62& 9.71& 2.90& 0.99 \\
\nodata&   UGC00168& 1& 5& 1.7$\pm$0.2& 8.80& 9.23& 2.57& 0.48 \\
\nodata&   UGC00179& 6& 5& 5.3$\pm$0.1& 9.30& 9.55& 2.44& 0.02 \\
IC1544&    UGC00204& 5& 5& 4.3$\pm$0.2& 9.22& 9.71& 2.73& 0.32 \\
IC1546&    \nodata & 5&15& 2.2$\pm$0.5& 8.91& 9.55& 2.45& 0.38 \\
\nodata&   UGC00228& 4& 5&12.8$\pm$0.1& 9.68& 9.63& 2.65&-0.23 \\
IC1552&    UGC00297& 5& 5& 2.5$\pm$0.2& 8.97& 9.63& 2.42& 0.30 \\
\enddata
\tablenotetext{1}{Asterisk implies the Arecibo observations were taken
  on the second observing run in October 2004 (see
  section~\ref{obs_optical}).} 
\end{deluxetable}

\clearpage
\begin{deluxetable}{cccccc}
\tablewidth{0pc}
\tablecaption{Cluster Properties\label{groups}}
\tablehead{
  \colhead{Cluster} &\colhead{$\%$(E+S0):\tablenotemark{1}}
  &\colhead{$\sigma$$_{v}$\tablenotemark{1}}
  &\colhead{X-ray Luminosity\tablenotemark{1}}
  &\colhead{electron density\tablenotemark{1}}
  &\colhead{$\rho$v$^{2}$} \\ 
  \colhead{Name} &\colhead{$\%$(S+IRR)}
  &\colhead{(km/s)} 
  &\colhead{(erg/s)}
  &\colhead{(cm$^{-3}$)}
  &\colhead{(km/s)$^{2}$ cm$^{-3}$} \\
}
\startdata
Coma       & 86:14$^{(a)}$ & 1010$^{(b)}$ & (0.5-3 Kev) 
25.7x10$^{43}$$^{(c)}$ & 2.5x10$^{-3}$$^{(d)}$ & 2550 \\  
Virgo      & 37:63$^{(e)}$ & 632$^{(f)}$  & (0.5-3 Kev)
4.4x10$^{43}$$^{(c)}$  & 6.4x10$^{-3}$$^{(d)}$ & 2556 \\  
Pegasus    & 18:82$^{(g)}$ & 240$^{(h)}$  & (0.2-4 Kev)
1.3x10$^{42}$$^{(i)}$ & 2x10$^{-4}$$^{(i)}$ & 12 \\ 
Eridanus   & 46:54$^{(e)}$ & 240$^{(e)}$  & (0.1-2 Kev)
2.5x10$^{41}$$^{(e)}$ & 2x10$^{-4}$$^{(e)}$ & 12 \\ 
Ursa Major & 15:85$^{(e)}$ & 150$^{(e)}$ & \nodata$^{(e)}$ 
& \nodata$^{(e)}$ & \nodata \\ 
\enddata
\tablenotetext{1}{(a)\citet{gh85}, (b)\citet{z90}, (c)\citet{jf78},
  (d)\citet{bs77}, (e)\citet{od05},
  (f)\citet{s01}, (g) measured by authors using the central group
  RA, DEC, and velocity constraints, 
  (h)\citet{rh82}, (i)\citet{c86}.}
\end{deluxetable}

\clearpage
\begin{deluxetable}{cccccc}
\tablewidth{0pc}
\tablecaption{HI Imaging\label{vla_offsets}}
\tablehead{
  \colhead{(1)}&\colhead{(2)}&\colhead{(3)}
  &\colhead{(4)}&\colhead{(5)}&\colhead{(6)} \\
  \colhead{Name}&\colhead{DEF}& \colhead{Displacement
    between}& 
  \colhead{Shift in HI center}& 
  \colhead{Asymmetry in HI}& \colhead{HI disk} \\
\colhead{}&\colhead{}& \colhead{HI and optical/D$_{o}$}
&\colhead{(yes or no)}&\colhead{(yes or no)}&\colhead{truncation}\\
}
\startdata
NGC7610    &-0.15  & 0.03 &no  &no &  0.3 \\
UGC12480   &-0.09  & 0.05 &no  &no &  0.1 \\
KUG2318+078& 0.20  & 0.09 &yes &yes& -0.5 \\
IC5309     & 0.29  & 0.02 &yes &yes& -0.5 \\
NGC7631    & 0.34  & 0.05 &no  &no & -0.2 \\
Z406-042   & 0.41  & 0.09 &no  &no & -0.4 \\
NGC7604    & 0.43  & 0.29 &no  &no & -0.2 \\
NGC7608    & 0.48  & 0.09 &yes &yes& -0.6 \\
NGC7615    & 0.85  & 0.08 &yes?&yes& -0.8 \\
\enddata
\end{deluxetable}

\end{document}